\documentclass[12 pt, amsmath,amssymb, aps, physrev]{revtex4-2}

\usepackage{graphicx}
\usepackage{dcolumn}
\usepackage{bm}
\usepackage{amsmath}
\usepackage{xcolor}

\newcommand{\sech}{\operatorname{sech}}

\begin{document}

\preprint{APS/123-QED}

\title{Braneworlds in Constant and Accelerated Motion and Their Causal Characteristics}%

\author{Ryan Debolt}
 \email{ryantdebolt@gmail.com}

\author{David Kagan}%
 \email{dkagan@umassd.edu}
\affiliation{%
 Department of Physics, University of Massachusetts Dartmouth\\
 285 Old Westport Road, Dartmouth, MA 02747, USA
}%


\date{\today}

\begin{abstract}
We generalize prior work on the signatures of bulk signals detected by brane-based observers in a spacetime with a compactified dimension. When such braneworlds move with constant velocities or constant proper accelerations in the extra dimension, the observers may witness apparent superluminal signaling. Our analysis includes tilted branes that are partially wrapped along the compact dimension. We identify parameters that help characterize various scenarios. Despite the apparent superluminality of bulk signals, we show that anisotropies in their propagation relative to the brane-based observer preserve causality. Some of the effects studied here could be the basis for alternative cosmological models, as well as observable signatures of braneworld scenarios.
\end{abstract}

\maketitle

\newpage

\section{Introduction}

Compactifying spatial dimensions breaks global Lorentz symmetry, leading to a preferred class of reference frames. The works \cite{PhysRevD.106.085001, PhysRevD.107.025016,POLYCHRONAKOS2023137917},  investigated the case of a spacetime with a single, circular extra dimension in the presence of branes that fill the remaining, uncompactified spatial dimensions. Observers confined to such space-filling branes are able to detect apparently superluminal signals if their branes move with constant velocity along the circular extra dimension. In some of these scenarios, the combination of observer motion along the brane and the brane's extra-dimensional motion leads to signals that appear to propagate backwards in time, but only for one-way trips. Nevertheless, analyses of the round-trip time for signals indicate that they cannot be used for actual faster-than-light communication. In other examples, wavefronts generated on the brane can be devised to propagate anisotropically, with source locations that exhibit an apparent drift, and other exotic-seeming signatures.

In this work, we develop a general framework that incorporates and extends the results described in  \cite{PhysRevD.106.085001, PhysRevD.107.025016,POLYCHRONAKOS2023137917}. We identify simple, key parameters that characterize the various scenarios, and explore situations involving accelerating branes and observers.

\section{The Preferred Frame}\label{sec:P-frame}

To set the scene, consider a bulk metric describing a flat, $d+2$ dimensional spacetime with a circular spatial direction,
\begin{equation}\label{metric}
ds^{2}_\textrm{bulk} = -dt^2 + dx^2+|d\bm{y}|^2 + dz^2, \ \ \ \  \bm{y} \in \mathbb{R}^{d-1}, \ \ \ \ z \approx z + 2\pi{R}.
\end{equation}
The coordinate $x$ is an arbitrarily chosen spatial axis, $\bm{y}$ refers to coordinates on a plane orthogonal to $x$, and $z$ is the extra compact dimension of radius $R$. In what follows, we will consider observers embedded in $d$-branes that fill $d$ of the spatial dimensions, with the full spacetime referred to as the `bulk.' The simplest configuration involves a $d$-brane that fills the dimensions corresponding to our $x$- and $\bm{y}$-coordinates, appearing as a point from the perspective of the circular $z$ direction. Later, we will consider other, more generic brane configurations.

The presence of a compact direction breaks the global Lorentz symmetry of spacetime, leaving a global $ISO(1,d)\times U(1)$ symmetry. Thus, there is a class of preferred frames that picks out both a special state of motion relative to the compact direction and a special orientation in the full bulk spacetime. A brane induces further spontaneous breaking of spacetime symmetries that, as discussed in \cite{PhysRevD.106.085001, PhysRevD.107.025016, POLYCHRONAKOS2023137917, kabat2023inducedlorentzviolationmoving}, allows for the possibility of apparently superluminal signal propagation. Consider a brane with an observer in a preferred frame. We let $z=0$ be the location of the brane in the extra dimension. The observer releases a flash of some massless bulk field, capable of propagating in all directions throughout the bulk at the speed of light. We will take the spacetime origin to be the source of the flash. Signals in the $z$ dimension periodically return to the origin due to the compactification, yielding an infinite series of images (referred to as `image charges') in the unwrapped $z$ dimension with initial preferred frame coordinates
\begin{equation}\label{preffered frame events}
t_n = x_n = \bm{y}_n = 0, \ \ \ z_n = 2\pi Rn \ \ \ n \in \mathbb{Z}.
\end{equation}
The light cones emanating from the image charges are described by the equation,
\begin{equation}\label{P-frame image charges}
(x-x_n)^2+ \vert \bm{y}-\bm{y}_n\vert^2 +(z-z_n)^2 = (t-t_n)^2.
\end{equation}
Setting $\bm{y} = 0$, we find that the light cone cross sections form circles in the $x-z$ plane,
\begin{equation}\label{P-frame image charges xz}
(x-x_n)^2+ (z-z_n)^2 = (t-t_n)^2.
\end{equation}
These circles lie along the $z$ axis in the frame and all have a radius of $t$. The circles sit in an envelope consisting of two tangent lines given by
\begin{equation}\label{p-frame envelope}
    x=\pm t.
\end{equation}
The observer first detects one of these image signals when the corresponding light cone intersects the $x$ axis. In the preferred frame, the first detection of each signal is at a time $\tau_{n,e}=\pm z_n$ and always occurs at the position $x=0$. 

\begin{figure}
    \centering
    \includegraphics[width=0.5\linewidth]{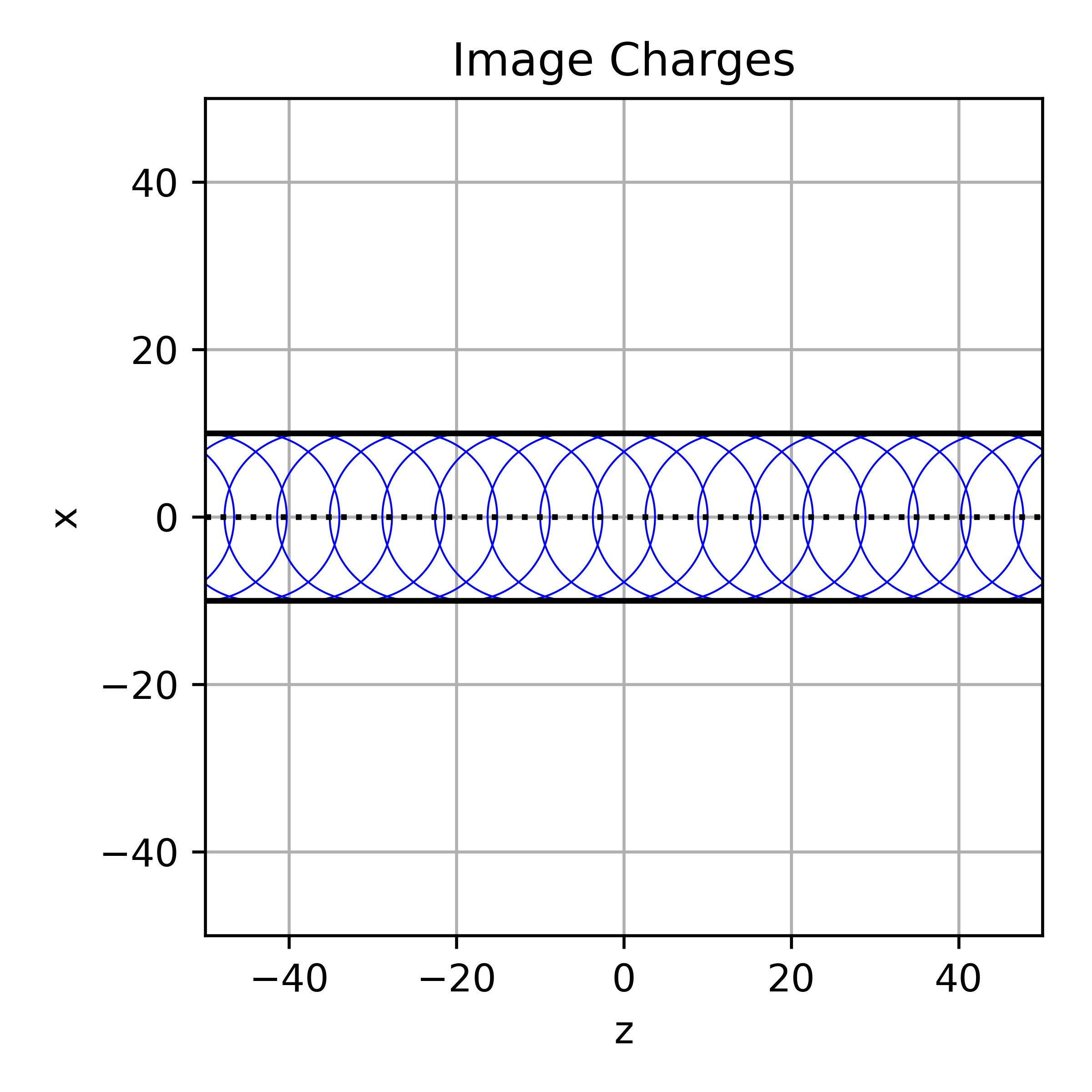}
    \caption[Preferred Frame Image Charges]%
    {Series of circles caused by the image charges in the preferred frame.}
    \label{fig:Preferred Frame Image Charges}
\end{figure}
As time passes, the wavefronts build up and form a cylinder-like front around the bulk's $z$ axis given by the equation
\begin{equation}\label{P-frame wavefront}
    x^2+\bm{y}^2=t^2.
\end{equation}
As expected, this wavefront propagates isotropically with a speed $c=1$. 
\begin{figure}
    \centering
    \includegraphics[width=0.5\linewidth]{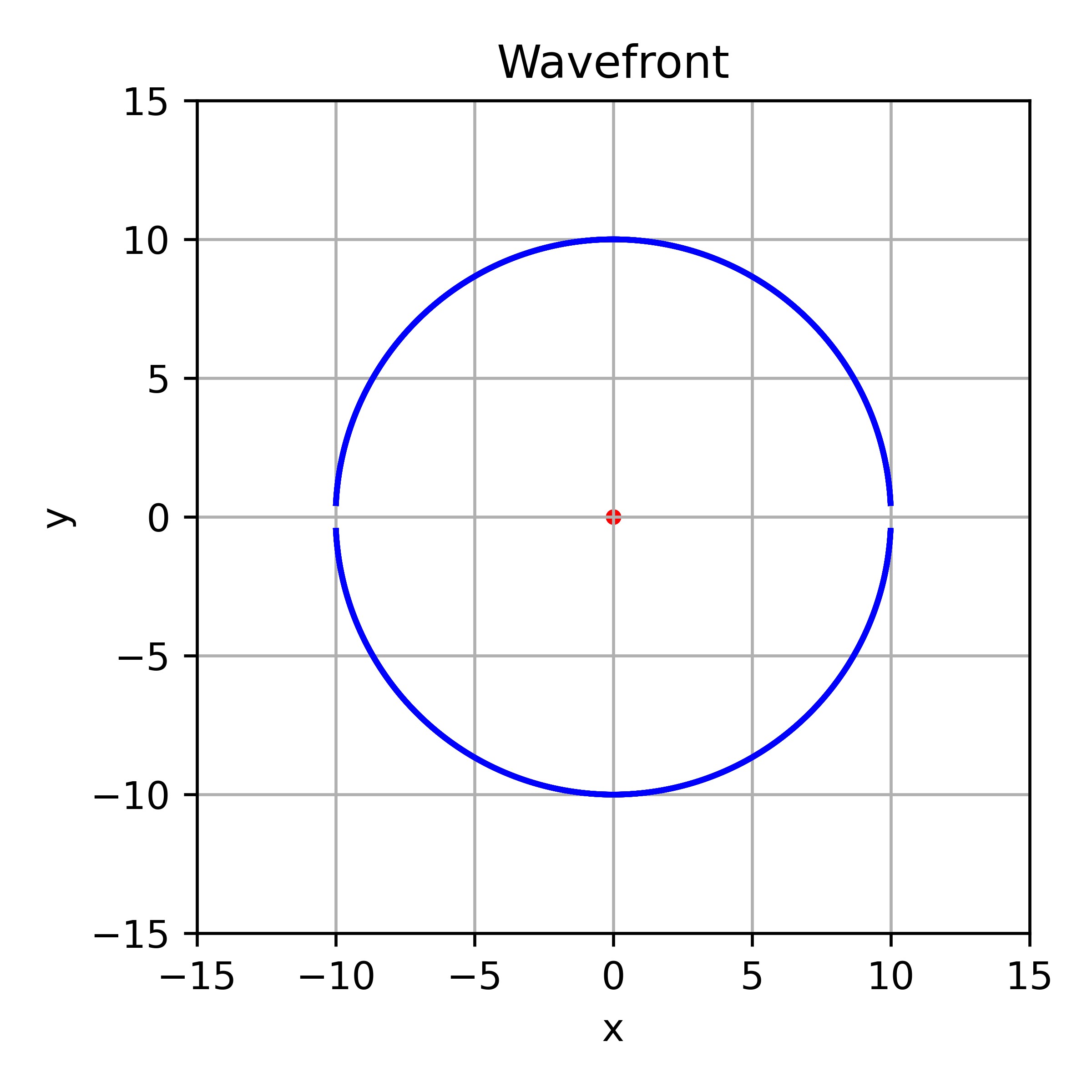}
    \caption[Preferred Frame Wavefront]%
    {Here we see the Isotropic wavefront generated on the Brane.}
    \label{fig:Preferred Frame Wavefront}
\end{figure}
Finally, let $\tau_n(x,\bm{y})$, denote the detection time of signals with winding number $n$ after they travel around the compact dimension as a function of the position $(x,\bm{y})$ of an observer on the brane,
\begin{equation}\label{P-frame detection times}
    \tau_n(x,\bm{y}) = -z_n \pm \sqrt{x^2+\bm{y}^2+z_n^2} .
\end{equation}
Here, we have substituted the brane's position in the extra dimension, $z=0$, into the expression (\ref{P-frame image charges} ). Note that for an observer at any given location $\left(x, \bm{y}\right)$ on the brane, the signals corresponding to $+n$ and $-n$ winding numbers arrive simultaneously, which is a characteristic of observing the signals from the preferred frame, but is not unique to this situation, as will be discussed later.

\begin{figure}
    \centering
    \includegraphics[width=0.5\linewidth]{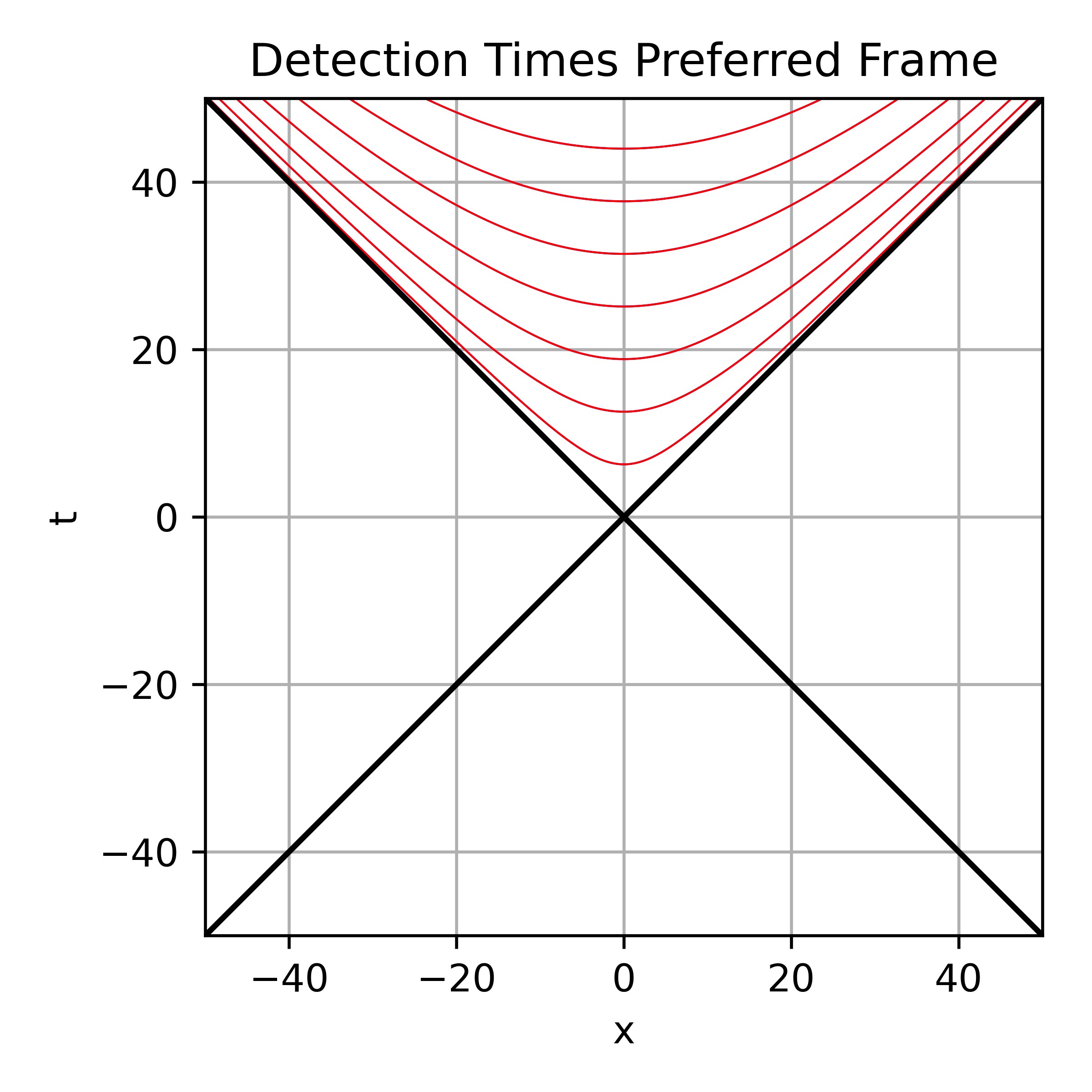}
    \caption[Preferred Frame Detection Times]{Detection times for each signal $n$ given as a function of position x.}
    \label{fig:Preferred Frame Detection Times}
\end{figure}

The detection time expression (\ref{P-frame detection times}) also demonstrates that observers at different locations on the brane disagree on the times the $|n|$ signals are detected, as can be seen in Figure \ref{fig:Preferred Frame Detection Times}. We note that the earliest detection time, $\tau_{n,e} = \tau_n(0,\bm{0})$, occurs at the coordinates $(0,\bm{0})$ corresponding to the observer at the brane's origin.
We thus conclude that the preferred frame is characterized by three key features: First, signals corresponding to the same magnitude winding number $|n|$, sent in opposing directions in the compactified dimension, must arrive simultaneously after circumnavigating the extra dimension according to all preferred-brane observers. Second, the observer located at the signal source, which we take to be the origin, must also be the first to detect each signal after the signals make their round-trip around the compact dimension. Third, the wavefront on the brane must propagate isotropically, with a center that remains static at the origin.

\section{Generalized Lorentz Branes}

Having established the features of the preferred frame, we turn to observations from the perspective of more general frames in which an observer is embedded on a potentially moving and ``tilted" brane. That is, the observer's brane may be partially wrapped around the compact dimension. The observer may also be moving along a given direction within the brane. We arrive at this observer's coordinates $t'', x'',$ and $z''$  by first considering the transformation to coordinates $t'$, $x'$, and $z'$, of an intermediate observer who moves along with the brane:
\begin{equation}\label{prime transformation matrix}
      \begin{bmatrix} 
        t{'} \\ x{'} \\  z{'}
    \end{bmatrix} = 
    \begin{bmatrix}
        \gamma & 0 & -\beta\gamma  \\
        0 & 1 & 0 \\
        -\beta\gamma & 0 & \gamma
    \end{bmatrix}
    \begin{bmatrix}
        1 & 0 & 0  \\
        0 & \cos\phi & \sin\phi \\
        0 & -\sin\phi & \cos\phi
    \end{bmatrix}
    \begin{bmatrix} 
        t \\ x \\  z
    \end{bmatrix},
\end{equation}
Here, the angle $\phi$ defines how tilted the brane's $x'$ axis is along the circular dimension. The parameter $\beta$ is the brane's velocity in the $z$ direction relative to the preferred frame, while $\gamma=1/(1-\beta^2)$. Writing out the intermediate transformations gives
\begin{equation}\label{prime transformations}
\begin{split}
    t{'} &= t\gamma + x\gamma\beta\sin\phi - z\gamma\beta\cos\phi, \\
    x{'} &= x\cos\phi + z\sin\phi,\\
    z{'} &= \gamma(-t\beta -x\sin\phi +z\cos\phi).
\end{split}
\end{equation}
Next, we boost to the moving observer's frame,
\begin{equation}\label{full transformation matrix}
\begin{split}
    \begin{bmatrix} 
        t{''} \\ x{''} \\  z{''}
    \end{bmatrix} &= 
    \begin{bmatrix}
        \Gamma & -B\Gamma & 0 \\
        -B\Gamma & \Gamma & 0 \\
        0 & 0 & 1
    \end{bmatrix}
    \begin{bmatrix} 
        t' \\ x' \\  z'
    \end{bmatrix}, \\
\end{split}
\end{equation}
where $B$ is the velocity of the brane-bound observer along the brane's $x'$ axis and $\Gamma = 1/(1-B^2)$.

The fully expanded transformations are
\begin{equation}\label{full transformations}
\begin{split}
    t{''} &= \Gamma(t\gamma + x(\gamma\beta\sin\phi - B\cos\phi) - z(\gamma\beta\cos\phi+B\sin\phi)), \\
    x{''} &= \Gamma(-tB\gamma -x(B\gamma\beta\sin\phi-\cos\phi) + z(B\gamma\beta\cos\phi+\sin\phi)),\\
    z{''} &= \gamma(-t\beta -x\sin\phi +z\cos\phi).
\end{split}
\end{equation}
These reduce to the well-known Lorentz transformations, as well as the transformations found in \cite{PhysRevD.106.085001, PhysRevLett.75.4724, POLYCHRONAKOS2023137917}, under the appropriate choices of parameter values. We simplify the form of our transformations by letting
\begin{equation}\label{parameters}
\begin{split}
        \mathcal{D} &= \Gamma\left(B\beta+\frac{1}{\gamma}\tan\phi\right) = \sinh(Z)\tanh(\zeta)+\cosh(Z)\sech(\zeta)\tan\phi, \\
        \mathcal{E} &= \Gamma\left(\beta+\frac{B}{\gamma}\tan\phi\right) = \cosh(Z)\tanh(\zeta)+\sinh(Z)\sech(\zeta)\tan\phi, \\
        \mathcal{F} &= \Gamma\left(B\beta-\frac{1}{\gamma}\cot\phi\right) = \sinh(Z)\tanh(\zeta)-\cosh(Z)\sech(\zeta)\cot\phi,  \\
        \mathcal{G} &= \Gamma\left(\beta-\frac{B}{\gamma}\cot\phi\right) = \cosh(Z)\tanh(\zeta) -\sinh(Z)\sech(\zeta)\cot\phi, \\
\end{split}
\end{equation}
where $\zeta$ is the rapidity associated with the brane's bulk motion and $Z$ is the rapidity associated with the observer's motion along the brane,
\begin{equation}\label{rapidities}
    \beta = \tanh(\zeta),\qquad B = \tanh(Z).
\end{equation}
Note that the preferred frame of Section \ref{sec:P-frame} corresponds to the case where $\mathcal{D}=\mathcal{E}=0$. We will describe the way that the characteristic features discussed in Section \ref{sec:P-frame} emerge as we explore the features of more general frames.

Our transformations can thus be condensed,
\begin{equation}\label{condensed matrix}
      \begin{bmatrix} 
        t{''} \\ x{''} \\  z{''}
    \end{bmatrix} = 
    \gamma
    \begin{bmatrix}
        \Gamma& \mathcal{G}\sin\phi & -\mathcal{E} \cos\phi\\
        -\Gamma B & - \mathcal{F} \sin\phi & \mathcal{D} \cos\phi \\
        -\beta & -\sin\phi & \cos\phi
    \end{bmatrix}
    \begin{bmatrix} 
        t \\ x \\  z
    \end{bmatrix},
\end{equation}
or in expanded form,
\begin{equation}\label{condensed transformations}
\begin{split}
    t{''} &= \gamma(t\Gamma + x\mathcal{G}\sin\phi - z\mathcal{E}\cos\phi), \\
    x{''} &= -\gamma(t\Gamma B +x \mathcal{F}\sin\phi - z\mathcal{D}\cos\phi), \\
    z{''} &= -\gamma(t\beta +x\sin\phi -z\cos\phi).
\end{split}
\end{equation}
The inverse transformations are given by
\begin{equation}\label{condensed inverse matrix}
       \begin{bmatrix} 
        t \\ x \\  z
    \end{bmatrix} = 
    \gamma
    \begin{bmatrix}
        \Gamma & \Gamma B  & \beta \\
        -\mathcal{G} \sin\phi & - \mathcal{F}  \sin\phi & -\sin\phi \\
       \mathcal{E}\cos\phi & \mathcal{D} \cos\phi & \cos\phi
    \end{bmatrix}
    \begin{bmatrix} 
        t'' \\ x'' \\  z''
    \end{bmatrix},
\end{equation}
with expanded form,
\begin{equation}\label{condensed inverse}
\begin{split}
    t &= \gamma(t{''}\Gamma + x{''}B\Gamma  + z{''}\beta), \\
    x &= -\gamma\sin\phi(t{''}\mathcal{G}+ x{''} \mathcal{F}+z{''}),\\
    z &= \gamma\cos\phi(t{''}\mathcal{E}+ x{''}\mathcal{D}+z{''}).
\end{split}
\end{equation}
Note that all of the $\bm y$ coordinates remain invariant under these transformations.

\subsection{Image charges}

The image charge coordinates in the boosted and tilted frame are
\begin{equation}\label{transformed events}
    \begin{split}
        z_n'' &= z_n\gamma\cos\phi, \\
        t_n'' &= -z_n''\mathcal{E},\\
        x_n'' &= z_n''\mathcal{D}.\\ 
    \end{split}
\end{equation}
Starting from equation \eqref{P-frame image charges},  we can describe the series of circles generated by using the inverse transformation laws we derived to transform the preferred frame coordinates in the equation, yielding:
\begin{equation}\label{transformed image charges}
\begin{split}
    &\gamma^2(t{''}\Gamma + x{''}B\Gamma  + z{''}\beta)^2 = \\ &|\mathbf{y{''}}|^2  + \left(-\gamma\sin\phi(t{''}\mathcal{G}+ x{''} \mathcal{F}+z{''})\right)^2 + \left(\gamma\cos\phi(t{''}\mathcal{E}+ x{''}\mathcal{D}+z{''})-z_n\right)^2.
\end{split}
\end{equation}
With a bit of algebra, this equation then simplifies to
\begin{equation}\label{condensed image charges}
\begin{split}
    (t{''}-t_n'')^2 &= (x{''}-x_n'')^2 + (z{''}-z_n'')^2 + |\mathbf{y}{''}|^2.
\end{split}
\end{equation}

\begin{figure}
    \centering
    \includegraphics[width=0.5\linewidth]{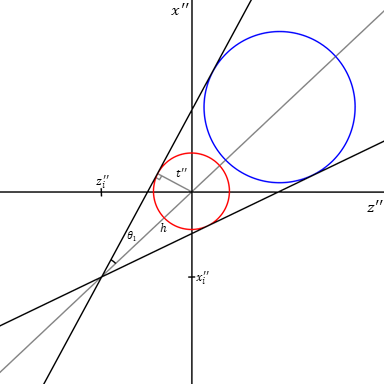}
    \caption[Envelope Drawing]{Here, we see the general construction of the series of circles and their envelope at some given time $t''$. }
    \label{fig:Envelope Drawing}
\end{figure}
For a slice of fixed $t{''}$, this equation represents a series of circles in the moving frame that correspond to the boundaries of the lightcones at that time, as pictured in \ref{fig:Envelope Drawing}. Observe that in the $x{''}-z{''}$ plane, the radius of the $n$th circular lightcone cross-section is
\begin{equation}\label{circleRadius-constV}
R_n=(t{''}+z_n''\mathcal{E}),   
\end{equation}
as given by \eqref{condensed image charges}. In the special case where the parameter $\mathcal{E}=0$, the radius of each circle is equal to the time in the frame, implying that the lightcones are all produced at the same time and expand at the same rate, as in the case of the preferred frame. However, the centers of the lightcones in a generic $\mathcal{E}=0$ frame differ from what would be observed in the preferred frame. Comparing equations \eqref{P-frame image charges} and \eqref{condensed image charges}, we note that the offsets on the $x''$ and $z''$ terms do not generally match those in the preferred frame. However, in all cases, the centers of the lightcones lie along a line, which we will refer to as the midline, that forms an angle $\theta_0$ with the $z{''}$ axis. The tangent of this angle gives the slope of the midline to be:
\begin{equation}\label{midline slope}
    m_{midline} = \tan(\theta_0) = \mathcal{D}.
\end{equation}
If the tilt angle $\phi$ of the brane is zero, then the midline reproduces the results found in \cite{PhysRevD.107.025016}. In the preferred frame we have $\mathcal{D}=0$, and thus all of the lightcone centers lie upon the $z$ axis. We recover that the lightcones' circular cross-sections all have radii equivalent to the preferred frame's time coordinate, and that they all lie along the $z$ axis.

Along with the midline, there is an envelope formed by two lines that run tangent to the series of circles, as seen in \ref{fig:Envelope Drawing}. In the preferred frame, these lines only intersect each other at the time $t=0$ when the envelope lines are identical to the midline. However, for any given time in a more general frame, the envelope and midline intersect at a point, forming a cone that is symmetric across the midline. The intersection point occurs where the radius of a circle centered at that point would be zero if we treat the winding number $n$ as continuous. For all cases, the coordinates of this point are
\begin{equation}\label{envelope intersections}
    \begin{split}
        x_i'' &= -\frac{t{''}\mathcal{D}}{\mathcal{E}}, \\
        z_i'' &= -\frac{t{''}}{\mathcal{E}}.
    \end{split}
\end{equation}  
Setting the brane's tilt to zero in our definitions of $\mathcal{D}$ and $\mathcal{E}$ in \eqref{parameters} allows us to recover the results of \cite{PhysRevD.106.085001, PhysRevLett.75.4724}. Let the angles $\pm\theta_1$ be the angles between the midline and the upper and lower lines of the envelope, respectively. To find $\theta_1$, note that the distance between the origin, where the $0$th signal is always centered, and the envelope's intersection point at time $t''$ is given by:
\begin{equation}\label{height}
\begin{split}
    h &= \sqrt{\left(-x_i''\right)^2 + \left(-z_i''\right)^2} \\
    & = t{''}\mathcal{E}^{-1}\sqrt{\left(1+\mathcal{D}^2\right)}.
\end{split}
\end{equation}
From (\ref{circleRadius-constV}), one finds that the circle formed at the origin always has a radius equal to the time in the frame, $R_0 =t{''}$. We can, therefore, form a right triangle using the line segment that connects the center of this circle with the point where the envelope intersects the circle, and the line segment that connects the center of this circle with the intersection point of the envelopes, as depicted in \ref{fig:Envelope Drawing}. We find that the sine of the angle formed by the envelope and midline is then,
\begin{equation}\label{theta1}
\begin{split}
    \sin(\theta_1) &= \mathcal{E}\left(\sqrt{\left(1+\mathcal{D}^2\right)}\right)^{-1} .
\end{split}
\end{equation}
The upper and lower envelope lines form angles $\theta_\pm = \theta_0\pm\theta_1$ with the $z{''}$, respectively. Taking the tangent of these angles yields the slopes,
\begin{equation}\label{envelope slopes}
\begin{split}
    m_{\pm} &= \tan(\theta_0\pm\theta_1) \\
    & = \frac{\mathcal{D}\sqrt{\mathcal{E}^{-2}(1+\mathcal{D}^2)-1}\pm1}{\sqrt{\mathcal{E}^{-2}(1+\mathcal{D}^2)-1}\mp \mathcal{D}} .
\end{split}
\end{equation}
Thus,
\begin{equation}\label{envelope equations}
    \left(x{''}-x_i''\right) = m_{\pm}\left(z{''}-z_i''\right),
\end{equation}
from which we find that at a time $t^"$, the envelopes cut the $x^"$ axis at the point
\begin{equation}\label{axis intersection}
\begin{split}
    x{''} &= t{''}\mathcal{E}^{-1}\left(m_{\pm} - \mathcal{D}\right).
    \end{split}
\end{equation}
The time derivative of (\ref{axis intersection}) yields the velocity of the envelope as it crosses the $x''$ axis,
\begin{equation}\label{envelope velocity}
\begin{split}
    v_{\pm} &= \mathcal{E}^{-1}\left(m_{\pm} - \mathcal{D}\right).
    \end{split}
\end{equation}
Substituting (\ref{envelope slopes}) yields
\begin{equation}\label{envelope velocity full}
\begin{split}
    v_{\pm} &= \pm\frac{\mathcal{E}^{-1}\left(1+\mathcal{D}^2\right)}{\sqrt{\mathcal{E}^{-2}\left(1+\mathcal{D}^2\right)-1}\mp \mathcal{D}}.
\end{split}
\end{equation}
As established in \cite{PhysRevD.107.025016}, these velocities can indeed exceed the speed of light. Furthermore, the velocities  (\ref{envelope velocity full}) diverge when $\mathcal{E}$ satisfies
\begin{equation}\label{critical point}
\begin{split}
    \mathcal{E}^2 &= 1.
\end{split}
\end{equation}
We refer to $\mathcal{E}=\pm 1$ as critical values, at which the signals in the frame appear to propagate instantaneously. When $|\mathcal{E}|>1$, the signals appear to propagate backward in time. Geometrically, at these critical values, either the upper or lower portion of the envelope runs parallel with the $x^"$ axis, corresponding to one of the enveloping lines developing a divergent slope.

\begin{figure}
    \centering
        \includegraphics[width=1\linewidth]{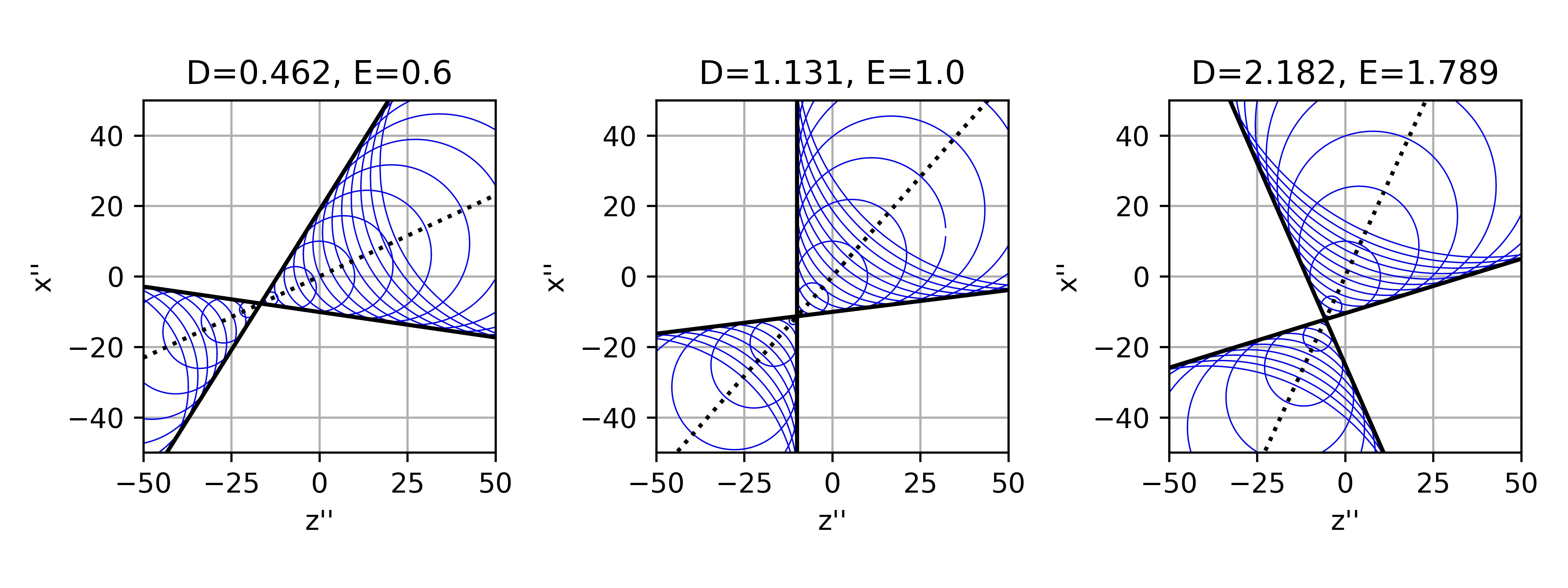}
        \caption[Lorentz Image Charges Criticality]%
        {These are examples of the series of circles in a subcritical, critical, and supercritical frame at time $t{''}=0$. Here, we can see how the values of $\mathcal{E}$  affect the propagation of the image charges in the $x''-z''$ plane.}
    \label{fig:Lorentz Image Charges Criticality}
\end{figure}

The parameter $\mathcal{E}$ thus allows us to classify our frames according to
\begin{equation}\label{criticalities}
\begin{split}
    |\mathcal{E}_{subcritical}| &<1, \\
    |\mathcal{E}_{critical}| &= 1, \\
    |\mathcal{E}_{supercritical}| &>1.
\end{split}
\end{equation}
Interpreted geometrically, these parameter regions are classified by the relationship between the $0$th circle and the intersection point of the envelope. For the envelope velocity to be well-defined and positive, corresponding to subcritical propagation, the intersection point must be greater in magnitude than the radius of the circle at the origin. The relationship $z_i''=-t''/\mathcal{E}$ requires that the magnitude $\mathcal{E}$ must therefore be less than one. Similarly, supercritical propagation corresponds to negative propagation speeds, which only occur when the intersection point is less than the magnitude of the radius. Figure \ref{fig:Lorentz Image Charges Criticality} depicts examples of each of these scenarios.
Note that when $\mathcal{E}$ is greater than $0$, the intersection point of the envelopes and midline is always located at a negative $x{''}$ coordinate. However, when $\mathcal{E}$ is less than $0$, this intersection point has a positive $x{''}$ coordinate. Furthermore, we find that when $\mathcal{E}=0$, the envelopes have the same slope as the midline, similar to their behavior in the preferred frame.

These results were found by geometrically analyzing lightcone cross-sections in the $x{''}-z{''}$ plane. We can cross-check our analysis by recalling that the envelope lines in the preferred frame are given by $x=\pm t$, and substituting this into our transformations. We find that the expression (\ref{envelope equations}) follows after some algebra.
\begin{figure}
    \centering
    \includegraphics[width=1\linewidth]{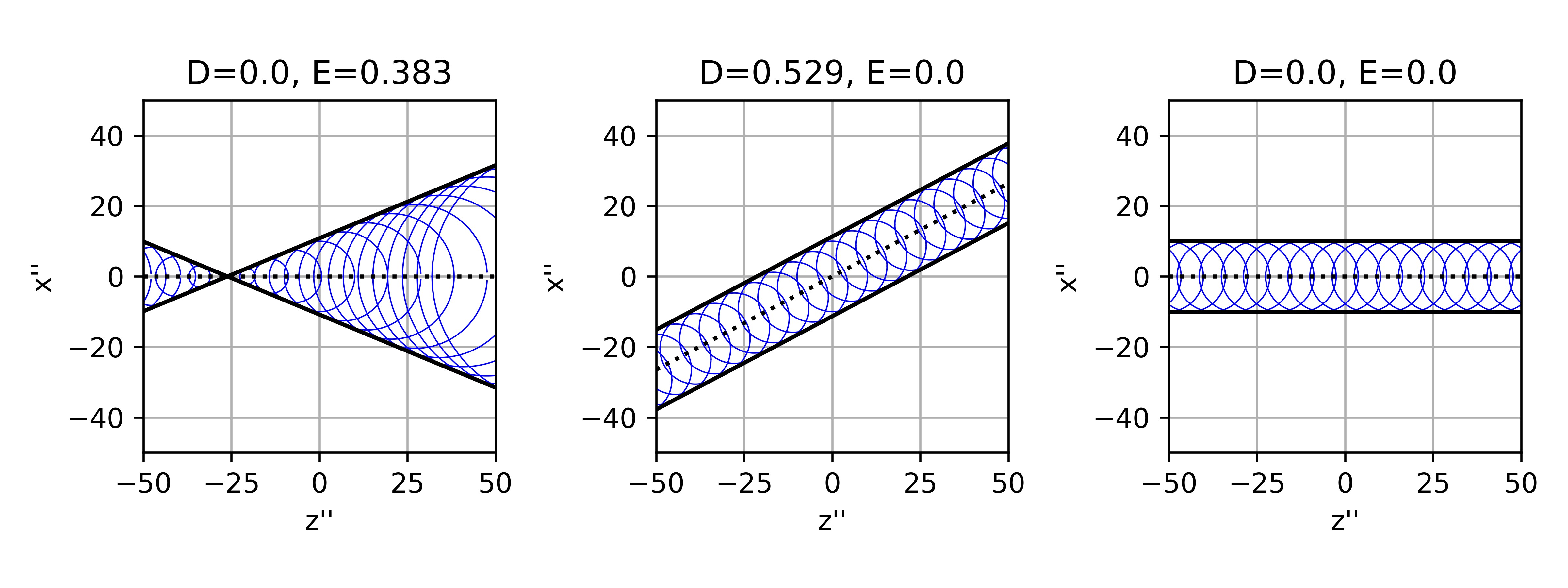}
    \caption[Lorentz Image Charges: Special Frames]%
    {Here we see how the series of circles generated by the image charges behave when special cases $\mathcal{D}=0$ and $\mathcal{E}=0$ are satisfied. When $\mathcal{D}$ and $\mathcal{E}$ are both zero, then we are describing the preferred frame.}
    \label{fig:Lorentz Image Charges: Special Frames}
\end{figure}

\subsection{Wavefronts}

Having examined characteristics of the tilted and boosted frame from the perspective of the $x''-z''$ plane, we now consider the wavefront observed on the moving brane. Starting with the preferred frame coordinates for a wavefront,  \eqref{P-frame wavefront} , and applying the inverse coordinate transformations \eqref{condensed inverse}, we find that the equation for the wavefront in the transformed frame is
\begin{equation}\label{transformed wavefront}
    \Gamma^2\gamma^2(t{''}+x{''}B)^2= \gamma^2\sin^2\phi(t{''}\mathcal{G}+x{''} \mathcal{F})^2+y{''}^2.
\end{equation}
Which can be re-expressed as,
\begin{equation}\label{alt wavefront}
    0 = \Gamma^2((Bt{''}+x{''})\cos\phi-\gamma\beta(t{''}+Bx{''})\sin\phi)^2+\mathbf{y{''}}^2 - \Gamma^2\gamma^2(t{''}+Bx{''})^2.
\end{equation}
As in \cite{POLYCHRONAKOS2023137917}, equation \eqref{alt wavefront} exhibits the anisotropy of the wavefront from the perspective of a moving observer on the moving brane, who witnesses different cross-sections of the lightcones in the $x''-\bm{y}''$ plane as the wavefronts intersect the brane. Assuming $\mathcal{E}<1$ for now, the wavefronts are elliptical with centers $x_c''$  at the point where $\frac{dy{''}}{dx{''}}=0$,
\begin{equation}\label{wavefront center}
    x_{c}'' = -t''\frac{\mathcal{G} \mathcal{F}\sin^2\phi-\Gamma^2B}{ \mathcal{F}^2\sin^2\phi-\Gamma^2B^2}.
\end{equation}
The center of these ellipses drift along the $x^"$ direction with a velocity
\begin{equation}\label{drift velocity}
     v_{c} = \frac{dx''_c}{dt''} = -\frac{\mathcal{G} \mathcal{F}\sin^2\phi-\Gamma^2B}{ \mathcal{F}^2\sin^2\phi-\Gamma^2B^2}.
\end{equation}
Transforming \eqref{alt wavefront} into polar coordinates $(r, \varphi)$ allows us to find the propagation speed $c_{\varphi}$ of the wavefront in any direction. First, let
\begin{equation}\label{sine parameters}
\begin{split}
     \mathcal{F}_s=&  \mathcal{F}\sin\phi, \\
    \mathcal{G}_s=& \mathcal{G}\sin\phi.
\end{split}
\end{equation}
The propagation speed is given by the time derivative of the radius of the wavefront. After some algebra, we find
\begin{equation}\label{propogation speed}
\begin{split}
    c_{\varphi} = \frac{ -\gamma^2\cos\varphi\left( \mathcal{F}_s\mathcal{G}_s-\Gamma^2B\right)\pm\gamma\sqrt{\Gamma^2\gamma^2\cos^2\varphi\left( \mathcal{F}_s-\mathcal{G}_sB\right)^2-\sin^2\varphi\left(\mathcal{G}_s^2-\Gamma^2\right)}}{\sin^2\varphi+\gamma^2\cos^2\varphi\left( \mathcal{F}_s^2-\Gamma^2B^2\right)}.
\end{split}
\end{equation}
The propagation speed in the $x{''}$ and $\mathbf{y}{''}$ directions are
\begin{equation}\label{directional propagation speeds}
\begin{split}
    c_{x{''}} &= \frac{-\left( \mathcal{F}_s\mathcal{G}_s-\Gamma^2B\right) \pm\Gamma\left( \mathcal{F}_s-\mathcal{G}_sB\right)}{\left( \mathcal{F}_s^2-\Gamma^2B^2\right)}, \\ 
    c_{\mathbf{y}{''}} &= \pm \gamma \sqrt{\Gamma^2-\mathcal{G}_s^2}.
\end{split}
\end{equation}
The propagation speeds have magnitudes greater than or equal to 1, again indicating apparent superluminality.

\begin{figure}
    \centering
    \includegraphics[width=1\linewidth]{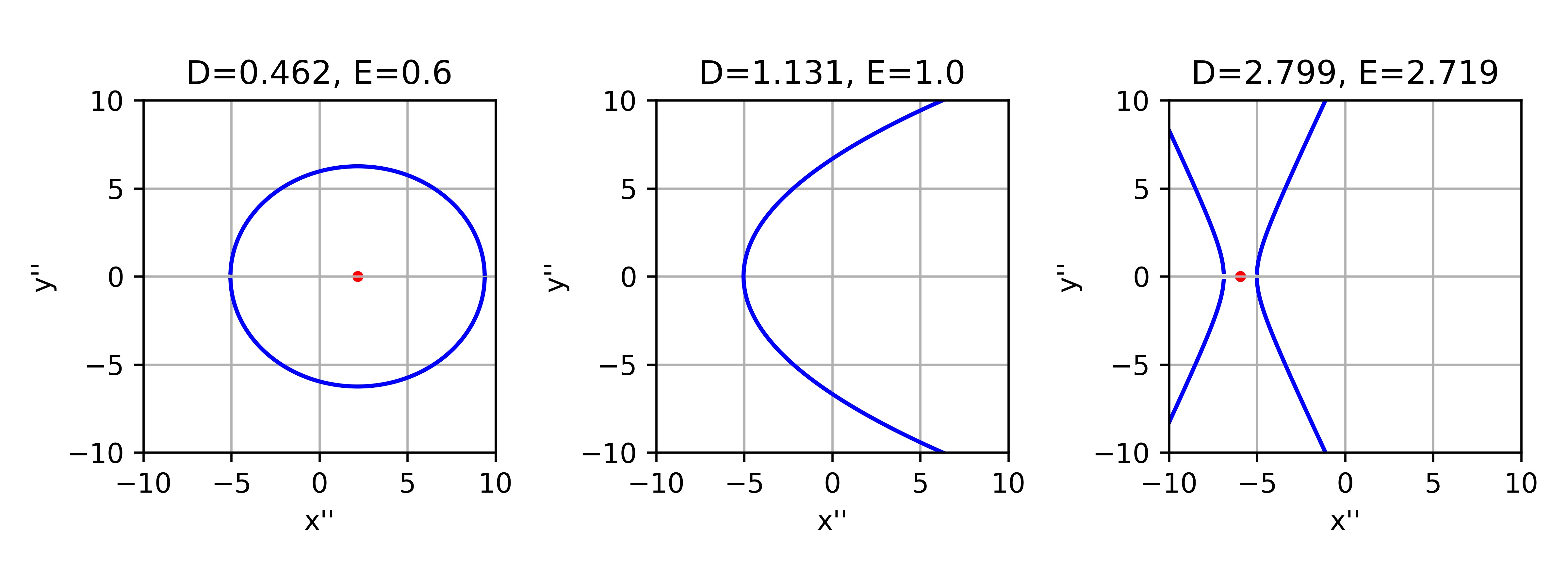}
    \caption[Lorentz Wavefront Criticalities]%
    {These plots demonstrate how the wavefront in the $x''-y'' $ frame evolves as the value for $\mathcal{E}$ changes. It also demonstrates how the perceived center of the wavefront drifts differently depending on this parameter.}
    \label{fig:Lorentz Wavefront Criticalities}
\end{figure}

We now want to observe the relationship between features of the wavefront and the parameters $\mathcal{D}$ and $\mathcal{E}$, particularly at special values. Note however that the wavefront features of interest are more directly dependent on $ \mathcal{F}$ and $\mathcal{G}$, defined in \eqref{parameters}. Nevertheless, these parameters' special values are tightly linked to those of $\mathcal{D}$ and $\mathcal{E}$. We find that at the critical point $\mathcal{E}=1$, the wavefront becomes parabolic, and the location of wavefront's center, \eqref{wavefront center}, and the center's velocity, \eqref{drift velocity}, diverge. When $\mathcal{E}>1$ the wavefront becomes hyperbolic, as seen in Figure \ref{fig:Lorentz Wavefront Criticalities}. The coordinate of the center, given by \eqref{wavefront center}, lies between the two branches of the hyperbola. The center's drift velocity also changes sign in this regime, and its magnitude is superluminal.

As the brane speed $\beta$ and observer speed $B$  together increase in magnitude, the magnitude of the position of the center and its drift speed \textit{decrease}. In the extreme case where both boosts approach unity, the center approaches a magnitude $|x_c''|=t{''}$  , and the drift speed approaches $|v_c|=1$. 
When $\mathcal{E}\to0$, the wavefronts can retain their ellipticity, but their center no longer drifts, meaning that $\mathcal{E}=0$ corresponds to the anisotropic driftless frames referred to as `tiltlike' in \cite{POLYCHRONAKOS2023137917}.

When the parameter $\mathcal{D}\to 0$, \eqref{transformed wavefront} implies that the wavefronts become circular, corresponding to an isotropic frame.\eqref{wavefront center} and \eqref{drift velocity} reduce to the case of driftless wavefront centers. However, the propagation speed \eqref{propogation speed} of these wavefronts has $c_\varphi \ge 1$, whereas the preferred frame requires that $c_\varphi =1$. Therefore, this result corresponds to the boostlike frames described in \cite{POLYCHRONAKOS2023137917}.

In the preferred frame, both $\mathcal{D}$ and $\mathcal{E}$ are zero, rendering the wavefronts driftless and isotropic. Wavefronts in the preferred frame must propagate at the speed of light. We conclude that the parameters $\mathcal{D}$ and $\mathcal{E}$ fully characterize the observed behavior of wavefronts in boosted, tilted frames with observers undergoing constant velocity motion. In summary,
\begin{equation}\label{driftless frames}
\begin{split}
    \mbox{Tiltlike (Driftless, Anisotropic):}&\ \ \mathcal{D}\ne0, \ \ \mathcal{E} = 0,\\
     \mbox{Boostlike (Driftless, Isotropic):}& \ \ \mathcal{D} = 0 ,\ \ \mathcal{E} \ne 0, \\
     \mbox{Preferred:}& \ \ \mathcal{D} = 0 ,\ \ \mathcal{E} = 0. \\
\end{split}
\end{equation}
\begin{figure}
    \centering
    \includegraphics[width=1\linewidth]{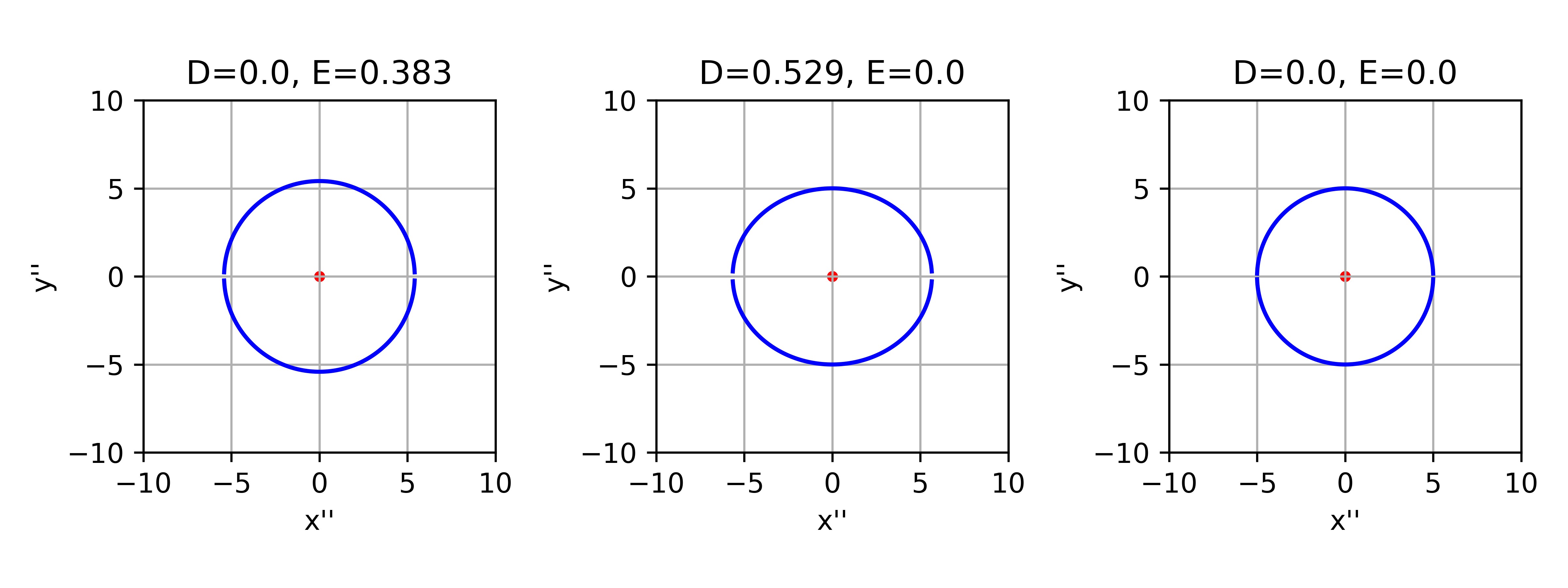}
    \caption[Lorentz Wavefront Special Frames]%
    {Here we see the wavefronts generated by our three special-case driftless frames. We can see that in these frames, the center never drifts from the origin. Frames where $\mathcal{D}=0$ generate a circular wavefront corresponding to isotropic propagation, while $\mathcal{E}=0$ frames do not, corresponding to anisotropic propagation.}
    \label{fig:Lorentz Wavefront Special Frames}
\end{figure}

\subsection{Detection Time}

The timing for bulk signal detection reveals important features of our brane-based observer's frame. Consider the detection time for a signal $n$ at any position $x{''}$. This is found by setting $z{''}=0$ in equation \eqref{condensed image charges} and yields
\begin{equation}\label{detection times}
\begin{split}
    \tau_n'' &=-z_n''\mathcal{E}\pm \sqrt{(x{''}-z_n''\mathcal{D})^2+{z_n''}^2+|\mathbf{y}''|^2}
\end{split}
\end{equation}
An observer located at the origin of this frame will detect these signals when they reach $x{''}=0$. Plugging this into the equation shows that this observer will first detect signals at:
\begin{equation}\label{origin detection times}
    \tau_n'' = -z_n''\left(\mathcal{E} \mp\sqrt{1+\mathcal{D}^2}\right)
\end{equation}
These detection times are always positive for positive $n$ signals. However, there may be earlier detection times for observers who are not located at the origin. Taking the $x''$-derivative of \eqref{origin detection times} and setting it to zero yields
\begin{equation}\label{detection time root}
    \frac{d\tau_n''}{dx{''}} = \frac{x{''}-z_n''\mathcal{D}}{\sqrt{(x{''}-z_n''\mathcal{D})^2+{z_n''}^2}}=0,
\end{equation}
giving the location of the earliest detection,
\begin{equation}\label{earliest detection position}
    x_{n,e}''=z_n''\mathcal{D}.
\end{equation}
On substitution into \eqref{detection times}, we find that the earliest detection time of a signal in the frame is
\begin{equation}\label{earliest detection times}
    \tau_{n,e}'' = \begin{cases} 
    -z_n''(\mathcal{E}-1), \quad n\geq0, \\
    -z_n''(\mathcal{E}+1 ), \quad n<0. \\
    \end{cases}
\end{equation}

When $\mathcal{D}=0$, the observer at the origin is the first to detect each signal. When $\mathcal{E}=0$, the earliest detection time satisfies $\tau_{n,e}''=z_n''$ independently of the signal winding number $n$. However, these detections do not necessarily occur at the same positions. Generally, the positive $n$ signals are first detected at a positive position $x''_e$, while the corresponding negative $n$ signals are detected at the same time at the corresponding position $-x''_e$. Thus, the observer at the origin does not generically make the earliest detections.

Furthermore, in cases where $\mathcal{E}=0$, the observer at the origin always detects signals with the same winding number magnitude simultaneously. Other observers may detect simultaneous signals, but these will not correspond to the matching winding number magnitudes. Thus, there are scenarios in which an observer at the origin would not detect apparent red or blue shifts in the signals and could be fooled into thinking they were in a preferred frame. In general, to determine whether or not they were in a preferred frame, an observer at the origin in this scenario would need information about detection times at different positions or information about the isotropy of the wavefront as discussed in the previous section. When both $\mathcal{D}$ and $\mathcal{E}$ are zero, the observer at the origin is the first to detect signals for all $n$ , as expected from our classification of the frames given in \eqref{driftless frames}.
\begin{figure}
    \centering
    \includegraphics[width=1\linewidth]{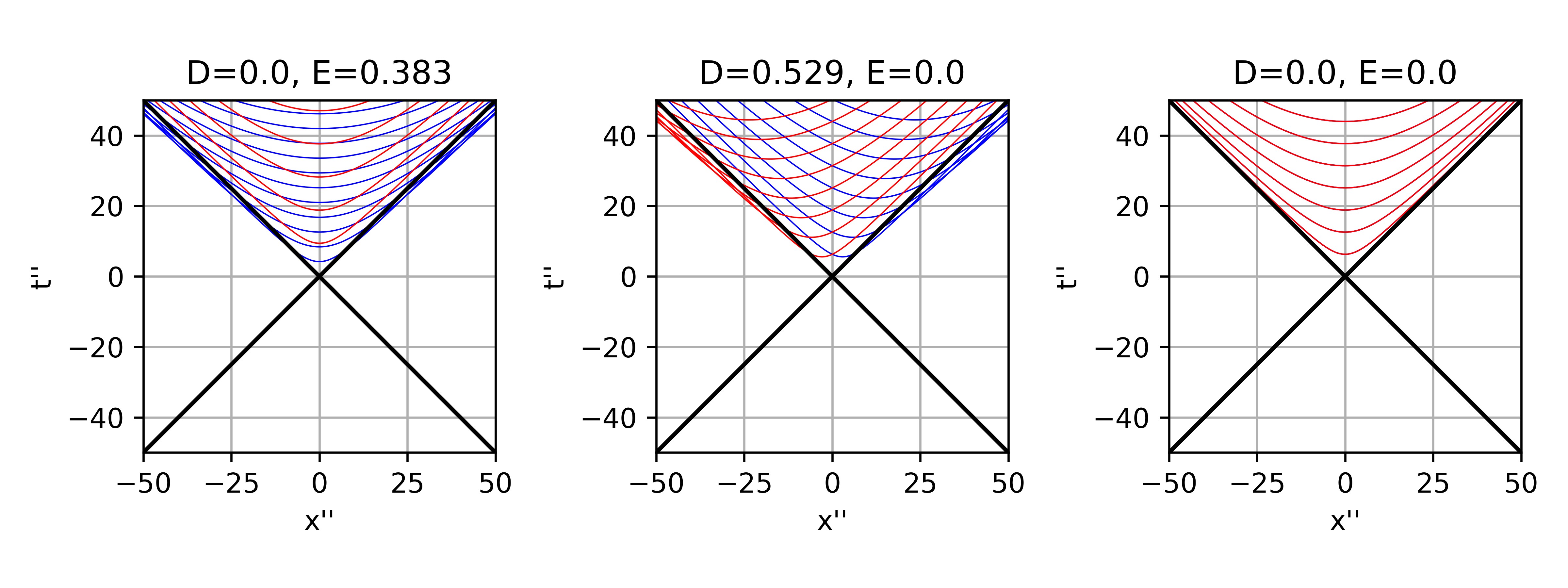}
    \caption[Detection Times Special Frames]%
    {Here we see the detection times in the  driftless frames. The blue curves represent signals that we would expect to be blueshifted, while red curves are redshifted, except in the special case $\mathcal{E}=0$ where positive and negative $n$ signals will appear to have the same frequency. We can see how in cases where the wavefront is isotropic, marked by when $\mathcal{D}=0$, then the observer at the origin will be the first to detect signals. However, in every other case, there will be an observer at a different position who can detect signals earlier.}
    \label{fig:Detection Times Special Frames}
\end{figure}

\subsection{Causality}

Observe that when $\vert \mathcal{E}\vert<1$, equation \eqref{earliest detection times} shows that detection times are positive for all $n$, meaning that in these frames, the signals are all detected after the original emission event at time zero. Recall, however, that $|\mathcal{E}|=1$ represents a critical value, beyond which some signals may appear to arrive from the detector's future, as can again be seen by substituting $|\mathcal{E}|>1$ into \eqref{earliest detection times}. Nevertheless, round-trip causal signaling is preserved in these frames, as we argue in what follows.

Note that in critical $\vert \mathcal{E}\vert = 1$ frames, the earliest detection times of signals are $\tau_{n,e}''=0$, but that either positive or negative $n$ can be seen to propagate instantly in these frames, but not both. If $\mathcal{E}=+1$, $n>0$ signals will propagate instantly, while negative signals will be first detected at $\tau_{n,e}''=2z_n''$. If $\mathcal{E}=-1$, then the opposite occurs.
\begin{figure}
    \centering
    \includegraphics[width=1\linewidth]{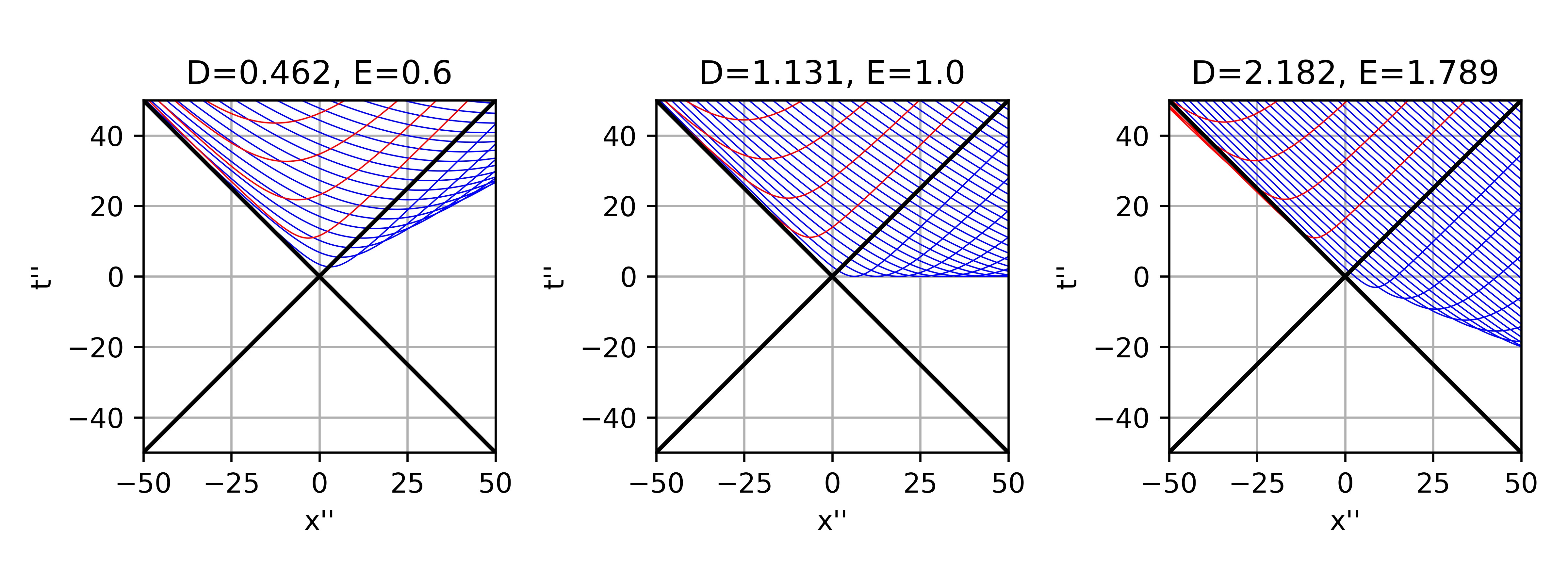}
    \caption[Lorentz Detection Times Criticalities]%
    {Here we see the detection times in scenarios spanning from subcritical to supercritical values of $\mathcal{E}$. As E becomes supercritical, we see how the detection times begin to become increasingly negative for the positive signals while negative signals remain positive.}
    \label{fig:Lorentz Detection Times Criticalities}
\end{figure}
As mentioned earlier, the seemingly backwards-in-time effect, also explored in \cite{PhysRevD.107.025016}, hints at the possibility of faster-than-light communication in such frames. However, this is not the case. Starting with
\begin{equation}\label{round trip times}
    T = l\left(\frac{1}{v_+}-\frac{1}{v_-}\right) = \frac{2l}{v_{eff}},
\end{equation}
we substitute the expressions for the envelope velocities \eqref{envelope velocity full}, yielding
\begin{equation}\label{full round trip times}
    T = 2l\sqrt{\frac{\mathcal{E}^{-2}\left(1+\mathcal{D}^2\right)-1}{\mathcal{E}^{-2}\left(1+\mathcal{D}^2\right)^2} }= \frac{2l}{v_{eff}},
\end{equation}
where the effective, round-trip velocity is given by
\begin{equation}\label{effective propagation velocity}
    v_{eff} = \sqrt{\frac{\mathcal{E}^{-2}\left(1+\mathcal{D}^2\right)^2}{\mathcal{E}^{-2}\left(1+\mathcal{D}^2\right)-1}}.
\end{equation}
The effective velocity's value is always greater than or equal to one, further demonstrating the apparent superluminality of the bulk signals in our frame. The round-trip time is thus bounded between $0<T \le 2l$. Thus, the bounds on the round-trip time first described in \cite{PhysRevD.107.025016} extend to our more general class of frames. We find that round-trip times are positive, and causality is preserved, but round-trip times can be made arbitrarily small by considering frames with large enough boost and tilt. 

\section{Constantly Accelerating Branes}

Having analyzed branes and observers related to the preferred frame by constant boosts, we now consider observers on branes undergoing constant proper acceleration. Starting from preferred frame coordinates $t,x,\bm{y},z$, we transform to the frame of an observer on the brane by applying K{\o}ttler-Moller transformations,
\begin{equation}\label{kottler moller}
\begin{split}
    t{'}&= \frac{1}{\alpha}\tanh^{-1}\left(\frac{t}{z+\frac{1}{\alpha}}\right), \\
    z{'}&=\sqrt{\left(z+\frac{1}{\alpha}\right)^2-t^2}-\frac{1}{\alpha}, \\
    t &=\left(z{'}+\frac{1}{\alpha}\right)\sinh(\alpha{t{'}}), \\
    z &=\left(z{'}+\frac{1}{\alpha}\right)\cosh(\alpha{t{'}}) -\frac{1}{\alpha}.
\end{split}
\end{equation}
Here, $\alpha$ denotes the brane's proper acceleration. These transformed coordinates are convenient because the accelerating observer and preferred observer are momentarily co-moving at the origin of this spacetime. 

\subsection{Zero Tilt Case}

\subsubsection{Image Charges}

Here, we consider a braneworld with no tilt and a frame that is not moving along the brane. The locations of the image charges generated are given by
\begin{equation}\label{rindler events}
    t_n' = \mathbf{x}_n' = 0, \ \ \ z_n' = 2\pi Rn \ \ \ n \in \mathbb{Z}.
\end{equation}
The light cones of these image charges generate a series of circles in the $\bm{x}'-z^{\prime}$ plane given by
\begin{equation}\label{rindler image charges}
    |\bm{x}'|^2 + \left(\left(z{'} +\frac{1}{\alpha}\right)\cosh(\alpha{t}{'})-\frac{1}{\alpha} - 2\pi{R}n\right)^2 = \left(\left(z{'} +\frac{1}{\alpha}\right)\sinh(\alpha{t}{'})\right)^2.
\end{equation}
Setting $u{'}= (z{'}+\frac{1}{\alpha})$  allows us to rewrite \eqref{rindler image charges} in a simpler form,
\begin{equation}\label{rindler image charges simple}
  \left(u_n'\sinh(\alpha t')\right)^2   = \left(u'-u_n'\cosh(\alpha t')\right)^2 +|\bm{x}{'}|^2.
\end{equation}
As in the non-accelerated cases, there is a midline and two lines that run tangent to this series of circles at any given time $t^{\prime}$. The opening angle of this envelope is related to the velocity of the brane along the extra dimension by,
\begin{equation}\label{beta function}
    \beta(t{'}) = \sin(\theta_{\pm}),
\end{equation}
with $\theta_{\pm}$ satisfying the relation
\begin{equation}\label{envelope tangent}
    \tan(\theta_{\pm}) = \frac{R_n}{u_n}.
\end{equation}
The radius of the $n$th circle, $R_n$, is equivalent to
\begin{equation}\label{circle radius}
    R_n = u_n'\sinh(\alpha t{'}).
\end{equation}
These results are consistent with the relation between the brane's velocity and its time-dependent rapidity $\zeta(t') = \alpha t'$ at time $t'$,
\begin{equation}\label{rapidity connection}
    \beta(t{'}) = \tanh\left(\alpha t{'}\right) = \mathcal{E}(t').
\end{equation}
The second equality follows from substituting the rapidity $\zeta=\alpha t'$, the tilt-free condition $\phi=0$ , and the frame's lack of motion along the accelerating brane $B=0$, into the definitions (\ref{parameters}) of our parameters $\mathcal{D},\mathcal{E}, \mathcal{F},$ and $\mathcal{G}$. We also have $\mathcal{D}(t')=0$.

The slopes of the envelopes are given by.
\begin{equation}\label{rindler envelope slope}
    m_\pm =  \pm\sinh(\alpha t{'}).
\end{equation}
The equation of the envelopes are determined by identifying the point where the radius of the $n$th circle tends toward zero. Using \eqref{circle radius}, we find that $R_n$ goes to zero at  $z{'} = -\frac{1}{\alpha}$, yielding the envelope equation
\begin{equation}\label{rindler envelope}
    x_\pm' = \pm u''\sinh(\alpha t{'}).
\end{equation}
\begin{figure}
    \centering
    \includegraphics[width=1\linewidth]{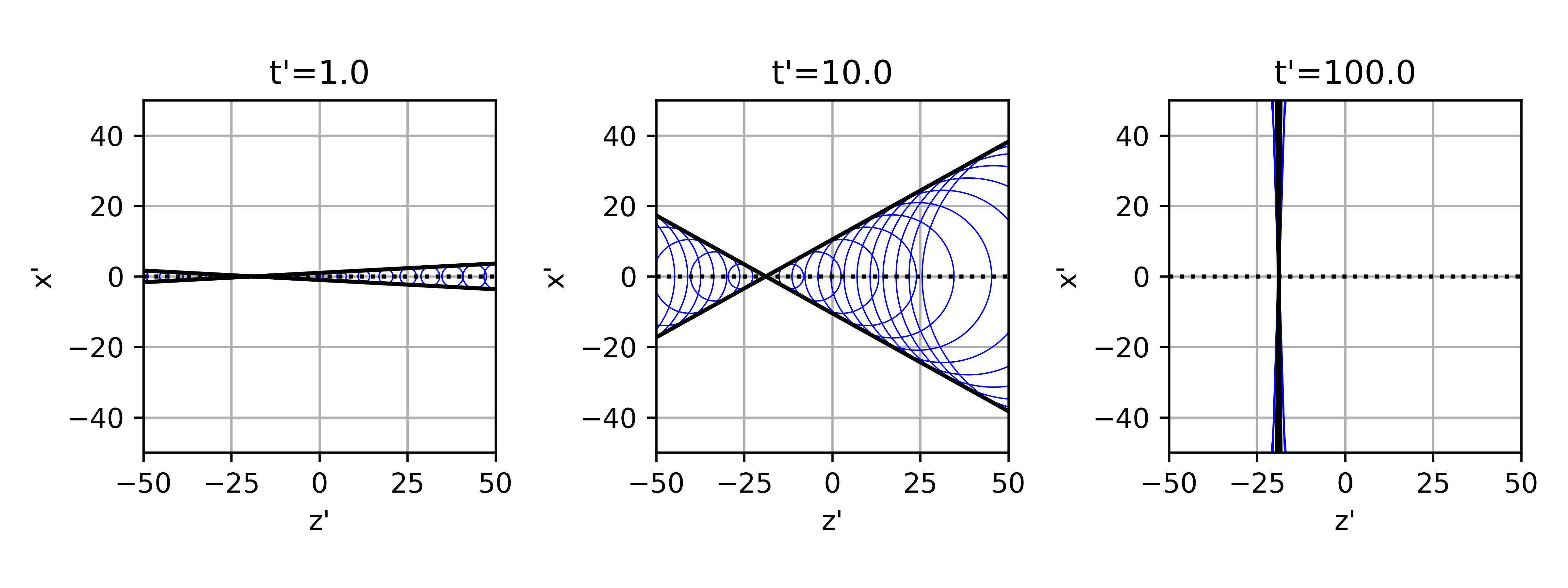}
    \caption[Rindler Image Charges]%
    {Rindler lightcones: These plots demonstrate how the lightcones, and the envelope running tangent to them, evolve with time. In particular, note how the origin of the envelope never moves from the Rindler horizon. Also note that as time goes to infinity, the envelope's slope diverges and the velocity of the brane approaches, but never exceeds the speed of light. These two facts make it impossible for the signals originating on the left side of the Rindler horizon to ever reach the observer located at the origin unless the motion were to change. } 
    s \label{fig:Rindler Image Charges}
\end{figure}
Unlike the case of a brane moving at constant velocity, the centers of the circles drift with time in the accelerated scenario with
\begin{equation}\label{circle center drifts}
    z_{nc}'=u_n'\cosh(\alpha t')-\frac{1}{\alpha}.
\end{equation}
For accelerating branes, the envelope's intersection point lies on the Rindler horizon, rendering its location independent of time, in contrast to cases with constant velocity. As such, signals cannot cross this boundary.  This is evident from \eqref{circle radius}, which shows that a circle centered at this intersection point has a fixed radius of zero and loses time dependence. The equation \eqref{circle center drifts}  also shows that the time-dependent term must disappear at this fixed point. The fixed nature of the envelope's intersection point prevents criticalities as the slope of the envelope cannot become divergent, and therefore signals cannot propagate instantaneously in these frames.

As time passes, \eqref{rapidity connection} shows that the brane's speed approaches the speed of light relative to the preferred frame. Initially, the envelope around the image charge signals expands rapidly, but eventually the growth of the envelope's angle begins to slow down. However, there is no upper limit on the radii of the image circles themselves, nor their rates of change,
\begin{equation}\label{rindler circle expansions}
    \begin{split}
        \dot{R_n} &= \alpha u_n'\cosh(\alpha t{'}).
    \end{split}
\end{equation}
As time increases, the propagation speed of the signals also increases without bound. Thus, circles corresponding to larger magnitude $n$ signals are always larger and grow faster than those of smaller magnitude $n$ signals. As a result, the larger $n$ circles tend to envelop the smaller ones. However, there is a region near the fixed point where the smaller $n$ circles cannot be enveloped. The existence of this region has implications for the causality in the system, as we discuss later when we consider detection times for these simple accelerated brane frames.

\subsubsection{Wavefronts}

The wavefront in the $x{'}-\mathbf{y}{'}$ subspace is given by
\begin{equation}\label{rindler wavefront}
    \bigg(\frac{1}{\alpha}\sinh(\alpha t{'})\bigg)^2=x'^2 +\mathbf{y}'^2.
\end{equation}
\begin{figure}[h]
    \centering
    \includegraphics[width=0.5\linewidth]{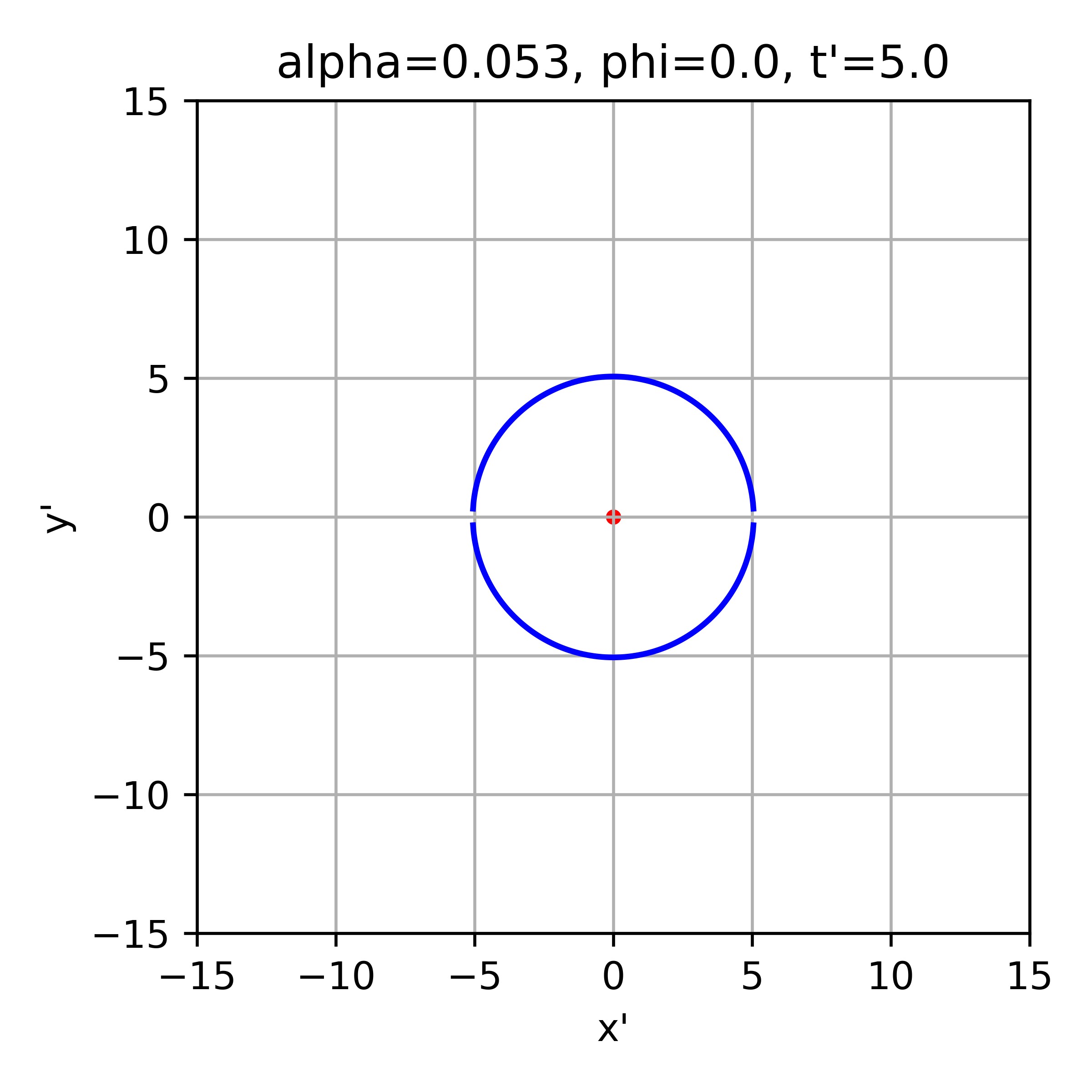}
    \caption[Rindler Wavefront]%
    {Rindler Wavefront: This demonstrates how the wavefront propagates. Note how the system is both isotropic and driftless. Connecting back to the constant velocity case, this process is continually boosting to and from frames where $\mathcal{D}=0$ as it accelerates. A tilt or additional boost will cause this wavefront to stretch or split as seen in the previous case.}
    \label{fig:Rindler Wavefront}
\end{figure}
The wavefront is always circular in this frame, with radius
\begin{equation}\label{rindler wavefront radii}
    R = \frac{1}{\alpha}\sinh(at{'}),
\end{equation}
and a propagation speed of
\begin{equation}\label{rindler propagation speed}
    c_\varphi = \cosh(\alpha t{'}).
\end{equation}
Thus, for $t'>0$ the wavefronts propagate superluminally and isotropically, naturally extending the results for Lorentz boosted branes explored in  \cite{PhysRevD.106.085001}. 

\subsubsection{Detection Times}

As in previous cases, round-trip signals cannot be used to communicate acausally. To demonstrate this, we solve for the detection times of the signals by an observer on the brane. Setting $z{'}=0$ in \eqref{rindler image charges simple} yields:
\begin{equation}\label{rindler detections}
    |\mathbf{x}{'}|^2  = -\frac{1}{\alpha^2} - u_n'^2 +\frac{2u_n'}{\alpha}\cosh(\alpha t{'}).
\end{equation}
\begin{figure}
    \centering
    \includegraphics[width=1\linewidth]{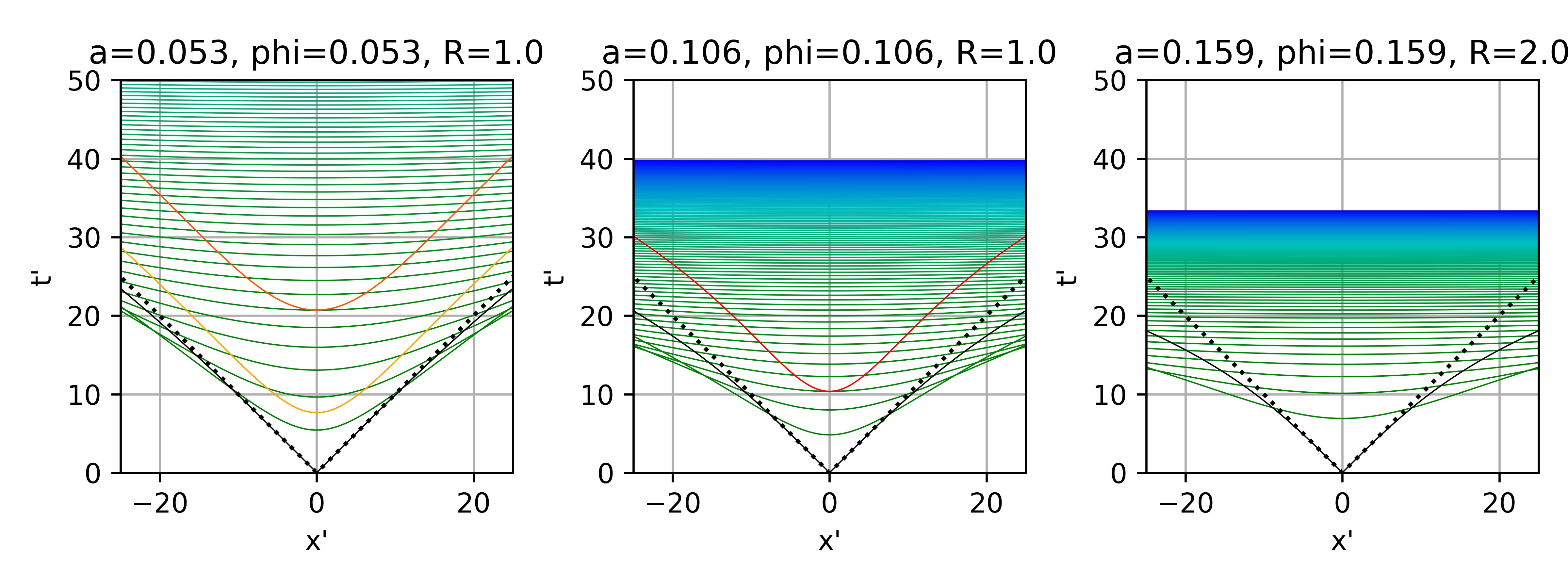}
    \caption[Rindler Detections]%
    {Rindler Detections: Here we see the detection time as a function of observer position for the first 100 signals. Positive signals in this frame continue to be blueshifted and increase in frequency indefinitely. Meanwhile, negative signals are increasingly redshifted until they disappear completely for signals originating beyond the Rindler horizon. Also note that as the magnitude of $x{'}$ increases, later $n$ signals are detected before earlier $n$ signals as a result of the flattening of the curve. The second plot demonstrates the case where $\alpha$ and $R$ are sufficiently large that no negative signals can be detected by the observer.}
    \label{fig:Rindler Detections}
\end{figure}
From \eqref{rindler detections}, the earliest detection occur at $\mathbf{x}{'} =0$, as these frames are isotropic. The detection times are given by
\begin{equation}\label{rindler earliest detections}
    \tau_{n,e}' = \pm\frac{1}{\alpha}\left|\ln(1+\alpha z_n)\right|.
\end{equation}
The earliest detection times of later $n$ signals are always greater than earlier signals, maintaining the causality of the system. However, observers located away from the origin may observe signals corresponding to a larger magnitude $n$ before the earlier signals. Furthermore, the existence of the Rindler horizon bounds the number of negative winding number signals by ensuring that all but a finite number remain behind the horizon relative to the brane, as illustrated in Figure \ref{fig:Rindler Detections}. These signals also appear to be increasingly redshifted relative to the observer until they disappear entirely.

In cases where the radius of the extra dimension, or the acceleration becomes sufficiently large in relation to each other, it becomes impossible for the observer to receive a signal from a negative $n$ source.

Positive $n$ signals are observed for all values of $n>0$. The signals are increasingly blueshifted as time goes on. Note that signals occurring before $t{'}=0$ are prevented from happening in these scenarios, as the chosen coordinates impose that the preferred frame and transformed frame agree at the origin, and that the velocity at that coordinate is zero.

\begin{figure}
    \centering
    \includegraphics[width=1\linewidth]{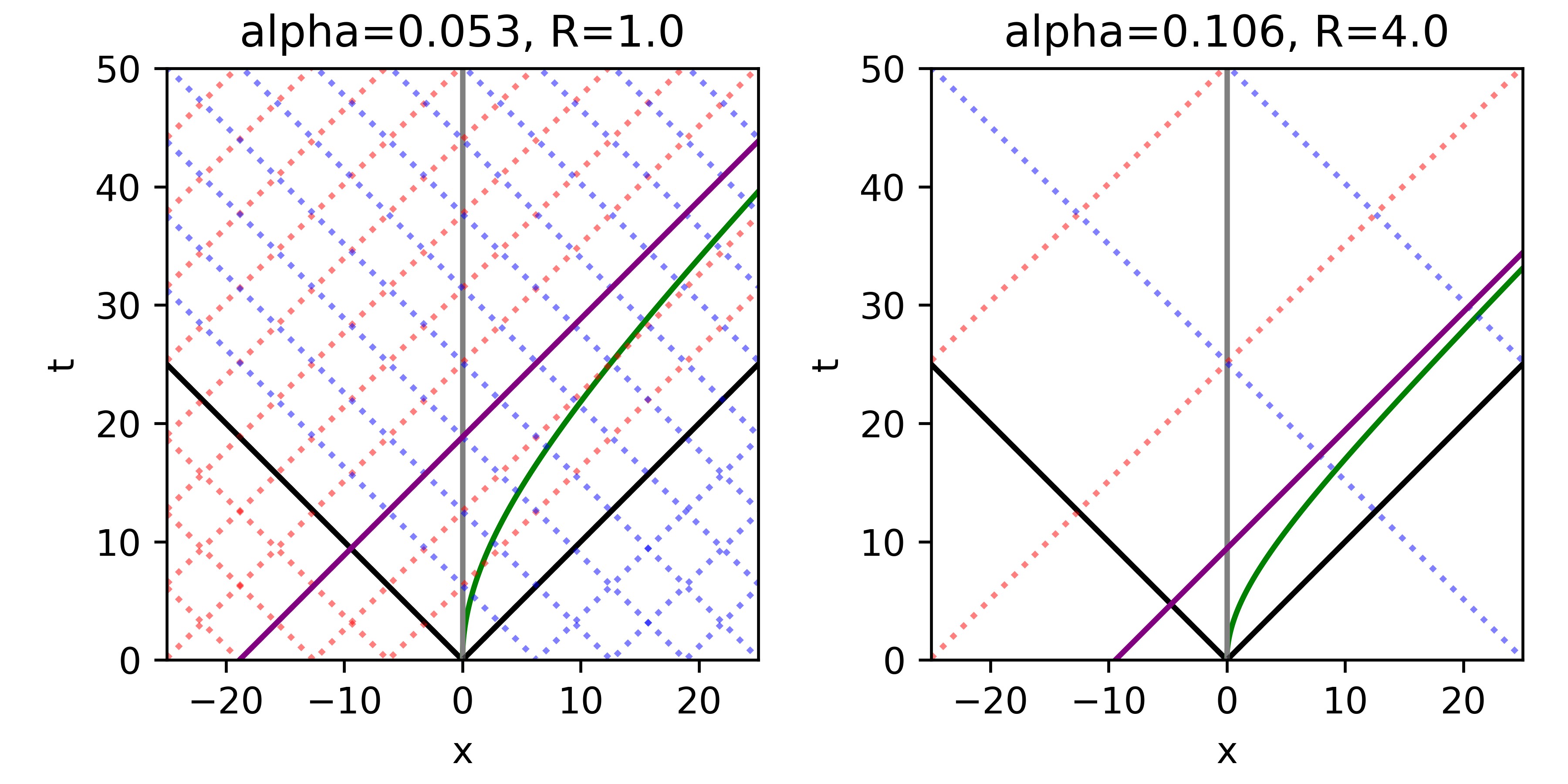}
    \caption[Worldline Intersections]%
    {Worldline Intersections: Here we see the path, shown by the green line, of the Rindler observer in the preferred frame. As this observer travels, they intersect with the lightcones of the frame, shown by the dotted lines. We see that as the position and time increase, the observer's worldline begins to asymptote to the Rindler horizon shown in purple. This serves to demonstrate how the signals that originate on the other side of the horizon will never interact with the accelerating observer. A Rindler horizon that originates before the first negative signal will prevent any negative signal from being observed. This occurs when $\alpha$ or $R$ become large as seen in the second plot.}
    \label{fig:Worldline Intersections}
\end{figure}

\subsection{Titled and Accelerated Brane} 

The transformations to a frame moving on an accelerating brane with a non-zero tilt are given by
\begin{equation}\label{tilted rindler transformation}
\begin{split}
    t{'} &= \frac{1}{\alpha}\tanh^{-1}\left(t\left(z\cos\phi-x\sin\phi+\frac{1}{\alpha}\right)^{-1}\right), \\
    x{'} &= x\cos\phi+z\sin\phi, \\
    z{'} &=\sqrt{\left(z\cos\phi-x\sin\phi+\frac{1}{\alpha}\right)^2-t^2} -\frac{1}{\alpha}.
\end{split}
\end{equation}
The inverse transformations are
\begin{equation}\label{tilted rindler inverse}
\begin{split}
    t &= u{'}\sinh\left(\alpha t'\right), \\
    x &= x'\cos\phi - \left(u{'}\cosh\left(\alpha t'\right)-\frac{1}{\alpha}\right)\sin\phi, \\
    z &= \left(u{'}\cosh\left(\alpha t'\right)-\frac{1}{\alpha}\right) \cos\phi+ x'\sin\phi,
\end{split}
\end{equation}
where
\begin{equation}
    \label{u''def}
    u' = z'+\frac{1}{\alpha}.
\end{equation}

Note that the rapidity function retains the same form as in the zero-tilt case, $\zeta=\alpha t'$. The parameters $\mathcal{D}$ and $\mathcal{E}$ are given by
\begin{equation}\label{tilted rindler parameters}
    \begin{split}
        \mathcal{D}&=\textrm{sech}(\alpha t')\tan\phi, \\
        \mathcal{E} &= \tanh(\alpha t').
    \end{split}
\end{equation}

\subsubsection{Image Charges}

The locations of the image charges for frames on a tilted accelerated brane are given by
\begin{equation}\label{tilted rindler events}
\begin{split}
    t_{n}' &= 0, \\
    x_{n}' &=  z_n\sin\phi, \\
    z_{n}' &=  z_n\cos\phi. \\
\end{split}
\end{equation}
Applying (\ref{tilted rindler inverse})  to \eqref{P-frame image charges} yields the formula for the series of circles in the tilted, accelerated frame,
\begin{equation}\label{tilted rindler image charges}
    \left(u_n'\sinh(\alpha t')\right)^2 =  \left(u'-u_n'\cosh(\alpha t')\right)^2+ \left(x'-x_n'\right)^2 + |\mathbf{y}'|^2.
\end{equation}
The Rindler horizon's location is as it was in the zero-tilt case. However, the midline develops a slope relative to the $z'$ axis, as shown by $\mathcal{D}\neq 0$ for all times. Note that $\mathcal{D}$ is indeed the slope of the midline in these scenarios. To see this, first note from our definition (\ref{u''def}) of $u''$ that the radius of the $i$th circle becomes zero at the Rindler horizon:
\begin{equation}\label{zeroeth circle}
\begin{split}
    z_{i}{'} &=-\frac{1}{\alpha}.
\end{split}
\end{equation}
Using (\ref{zeroeth circle}) and  \eqref{tilted rindler events} we find
\begin{equation}\label{x envelope intersection}
    x_i'=-\frac{\tan\phi}{\alpha}.
\end{equation}
The 0th circle is unaffected by rotations, as the rotation dependency is tied to $n$. Using  \eqref{circle center drifts} we set the center of the 0th circle coordinate to lie on the $z'$ axis at the coordinate:
\begin{equation}\label{tilted rindler circle centers}
\begin{split}
    z_{c,0}'&=\frac{1}{\alpha}\cosh(\alpha t')-\frac{1}{\alpha} \\
    x_{c,0}' &= 0
\end{split}
\end{equation}
\begin{figure}
    \centering
    \includegraphics[width=1\linewidth]{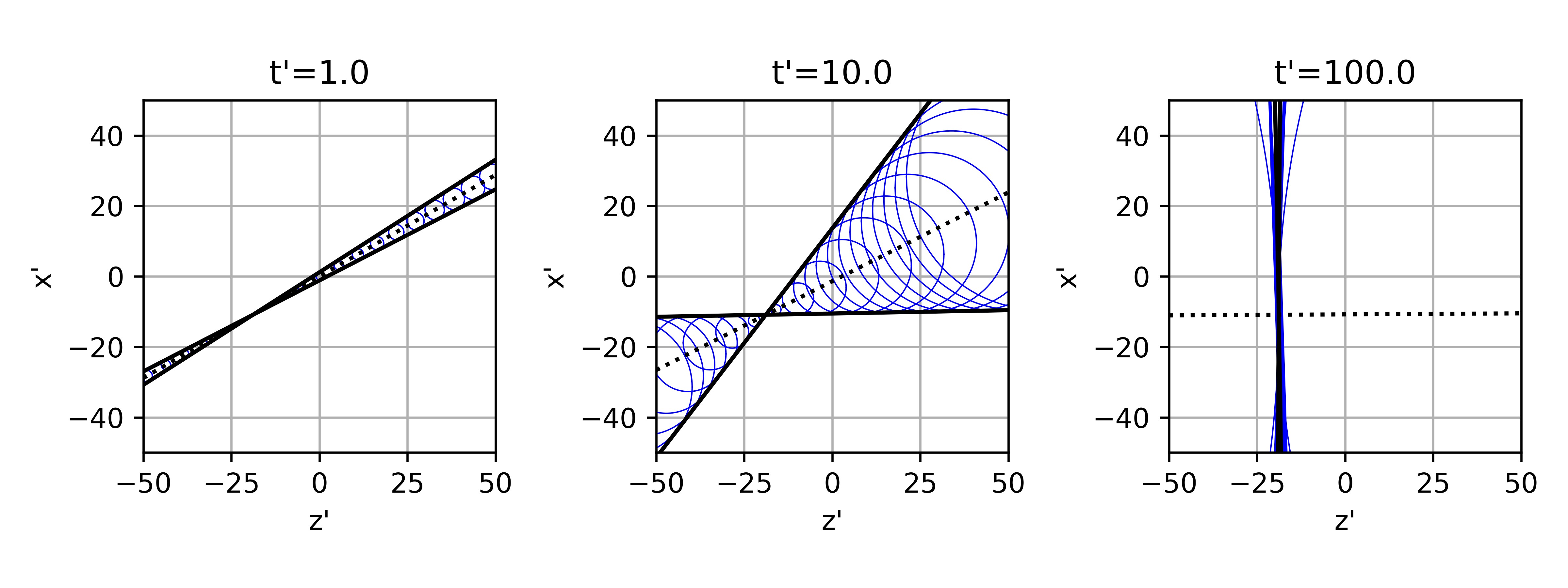}
    \caption[Tilted Rindler Lightcones]%
    {Tilted Rindler Lightcones: Here we see the behavior of the lightcone circles as the time increases. We see that there is an additional drift in $x'$ along with the drift seen in $z'$ demonstrated in the simple case. This results in the midline decreasing in slope over time as it begins to approach the $z'$ axis. }
    \label{fig:Tilted Rindler Lightcones}
\end{figure}
The slope of the midline follows from the locations of the two points with coordinates given by $(z_i',x_i')$ and $(z_{c,0}',x_{c,0}')$. Some trigonometry yields
\begin{equation}\label{tilted rindler slope}
    m_{midline} = \frac{\tan\phi}{\cosh(\alpha t')} = \mathcal{D},
\end{equation}
agreeing with \eqref{tilted rindler parameters}.

Substituting the inverse transformations \eqref{tilted rindler inverse} into our expression for the envelope lines in the preferred frame \eqref{p-frame envelope} yields the equation for the envelopes in the tilted, accelerating frame:
\begin{equation}\label{tilted rindler envelopes}
    \left(x'+\frac{\tan\phi}{a}\right)=m_\pm u',
\end{equation}
where $m_\pm$ is once again given by  \eqref{envelope slopes} and is now time-dependent.

As in tilt-free cases, the propagation speed only asymptotically approaches $\mathcal{E}=1$, the critical point described in the constant velocity cases. An observer at the origin never sees backward-in-time signals arising when observer's brane accelerates from rest at $t'=0$. This is due to such an observer's inability to see image charges and envelopes cross the Rindler horizon, thereby preventing the envelope's slope from changing sign. This is consistent with the corresponding constant velocity case when the observer is stationary with respect to the brane. In those constant velocity frames, the magnitude of $\mathcal{E}$ cannot reach or exceed $1$ . We expect that if the observer on the brane were moving, and the brane were boosted from a critical or supercritical frame, then we would see instantaneous and backward-in-time signaling. However, as noted in Figure \ref{fig:Rindler Detections}, observers at locations other than the origin can begin to see late $n$ signals arrive before earlier $n$ signals due to the circles for the larger $n$ image charges growing faster than the small $n$ charges and enveloping them. 

\subsubsection{Wavefronts}

The wavefront along the brane in this frame is given by
\begin{equation}\label{tilted rindler wavefront}
    \bigg(\frac{1}{\alpha}\sinh(at{'})\bigg)^2 = \bigg(x'\cos\phi-\frac{1}{\alpha}\left(\cosh(\alpha t')-1\right)\sin\phi\bigg)^2 +{\mathbf{y}}'^2.
\end{equation}
We noted earlier that we can never reach criticality in this scenario. Therefore, wavefronts must always be ellipses in this frame. Furthermore, since $\mathcal{D}$ is not necessarily zero in this case, the system will be anisotropic and have a drift. The center of the wavefront is then shown to be given by:
\begin{equation}\label{tilted rindler wavefront center}
    x_c'=\frac{1}{\alpha}(\cosh(\alpha t')-1)\tan\phi
\end{equation}
and drifts with velocity:
\begin{equation}\label{tilted rindler drift velocity}
    v_c=\sinh(\alpha t')\tan\phi
\end{equation}
\begin{figure}
    \centering
    \includegraphics[width=0.5\linewidth]{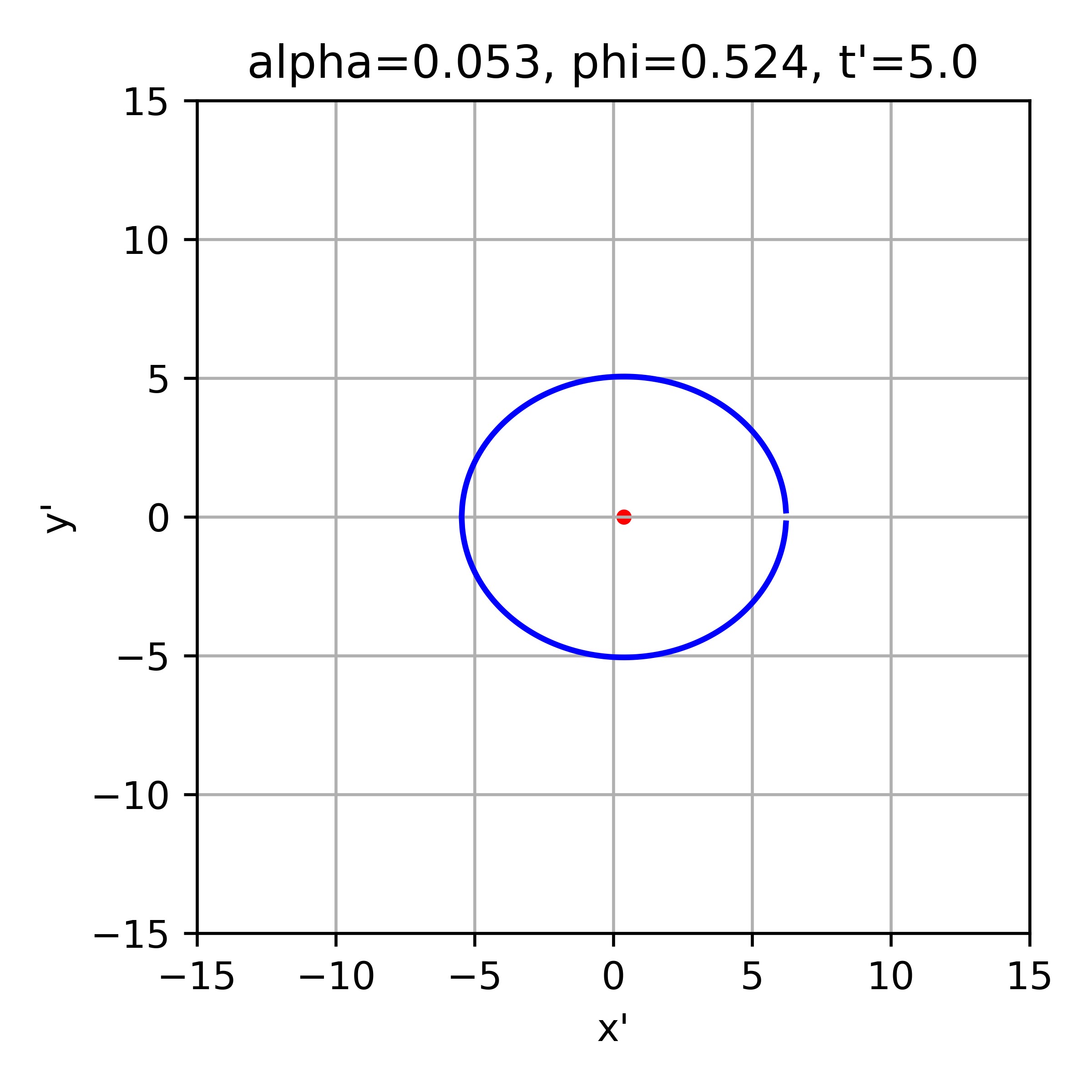}
    \caption[Tilted Rindler Wavefront]%
    {Tilted Rindler Wavefront: Here we see the ellipsoid generated at a time $t'=5$ in the transformed frame. Unlike the Lorentz cases, the rate of expansion of the wavefront is not constant. In fact, this growth, and subsequently the propagation speed of light, becomes incredibly rapid the later the time. However, they never become hyperboloids as the criticality condition is never met without the introduction of a secondary boost.}
    \label{fig:Tilted Rindler Wavefront}
\end{figure}
The propagation speed in a direction $\varphi$ in this case then becomes:
\begin{equation}\label{tileted rindler propagation speed}
    \begin{split}
        c_{\varphi}=& \sinh(\alpha t')\left(2\left(\cos^2\phi\cos^2\varphi+\sin^2\phi\right)\right)^{-1} \\ 
        &  \times \Biggr[\left(\cos(\varphi)\sin(2\phi)\right) + \left(\sinh^2\left(\frac{\alpha t'}{2}\right)\left(1+\cos^2(\phi) \cosh(\alpha t') - \cos(2\varphi)\sin^2(\phi)\right)\right)^{-1/2} \\
        &\times\Biggr(\sqrt{2}\left(\cos^2(\phi)\cosh(\alpha t')+\sin^2(\varphi)\sin^2(\phi)\right) \\ 
        &  \pm \sqrt{\sinh^2(\alpha t')\left(\cos^2(\varphi)\cos^2(\phi)+\sin^2(\phi)\right) - 4\sin^2(\varphi)\sin^2(\phi)\sinh^4\left(\frac{\alpha t'}{2}\right) } \Biggr)
        \Biggr]
    \end{split}
\end{equation}
As in previous cases, the propagation speed is always greater than or equal to unity in all directions. We also note that as time approaches infinity, the value of $\mathcal{D}$ begins to approach zero for all cases except $\phi=\pi/2$. This would imply that even an initially highly tilted braneworld would appear nearly isotropic if accelerated for a long enough time. However, the drift remains in this case as the center and drift velocity both approach infinity. 

\subsubsection{Detection times}

The detection time of each signal as a function of $x''$ is given by
\begin{equation}\label{tilted rindler detection times}
    \tau_n' = \frac{1}{\alpha}\cosh^{-1}\left(\frac{\alpha}{2u_n'}\left((x'-x_n')^2+u_n'^2+\frac{1}{\alpha^2}\right)\right).
\end{equation}
The observer at the origin will then first detect signals at time
\begin{equation}\label{tilted rindler origin detections}
    \tau_n' = \frac{1}{\alpha}\cosh^{-1}\left(\frac{\alpha}{2u_n'}\left(x_n'^2+u_n'^2+\frac{1}{\alpha^2}\right)\right).
\end{equation}
The earliest detection times will occur at time
\begin{equation}\label{tilted rindler earliest detections}
    \tau_{n,e}' = \frac{1}{\alpha}\cosh^{-1}\left(\frac{\alpha}{2u_n'}\left(u_n'^2+\frac{1}{\alpha^2}\right)\right).
\end{equation}
\begin{figure}
    \centering
    \includegraphics[width=1\linewidth]{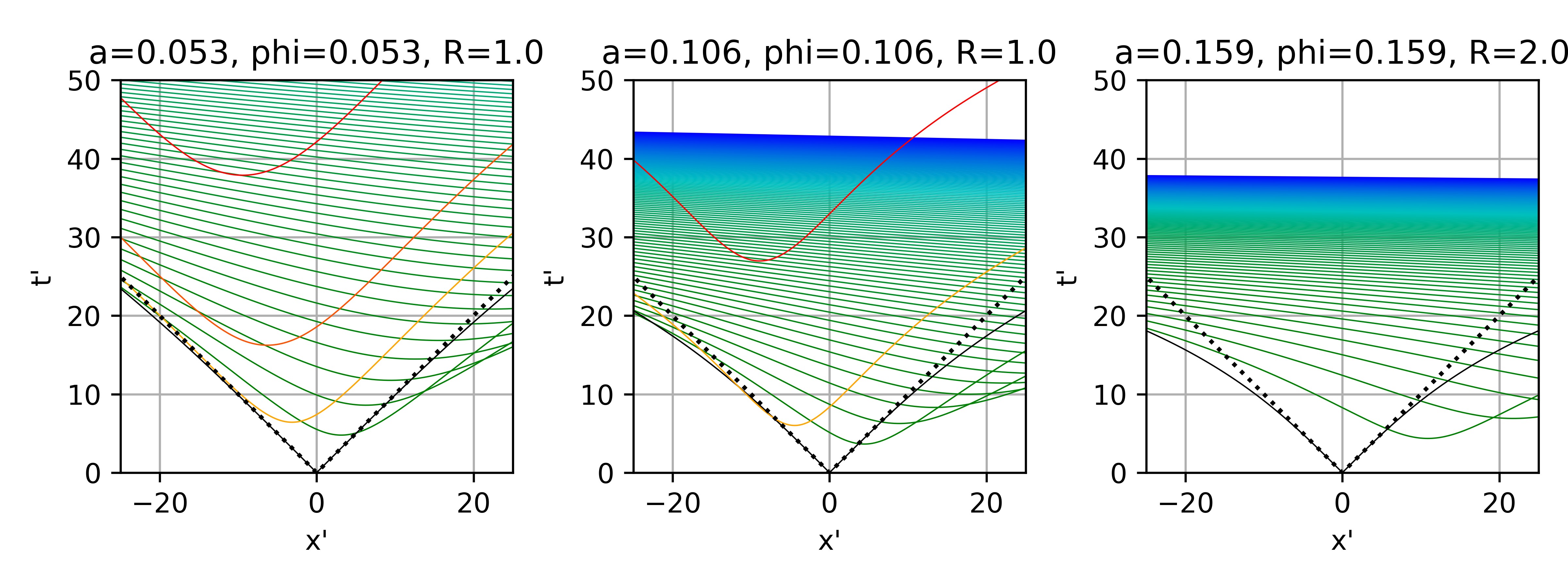}
    \caption[Tilted Rindler Detections]%
    {Tilted Rindler Detections: Here we see how the tilt affects the detections of signals in the frame. In particular, a set of parameters that would normally prevent negative signal detection can now see evidence of these signals when a tilt is introduced.}
    \label{fig:Tilted Rindler Detections}
\end{figure}
Thus, there is a limit on the number of negative signals that the observer will detect. However, the number of negative signal detections now depends on $\phi$. When $\phi$ approaches $\pi/2$, the number of detectable negative $n$ signals diverges. As in the non-tilted, accelerating case, there are no detections that occur before emission according to these observers. Furthermore, the earliest detections of $n$ signals maintain their order. We therefore conclude that causality is maintained in tilted, accelerated frames. However, as stated earlier, there will be observers at different positions who detect signals out of order due as seen from the flattening of the detection curve in \ref{fig:Tilted Rindler Detections}.

\section{Conclusions}

This work extends that of \cite{PhysRevD.106.085001, PhysRevD.107.025016, POLYCHRONAKOS2023137917}. We consider more general scenarios involving a brane traveling at constant velocity and also investigate accelerating branes. We identify parameters that are useful for classifying scenarios that differ in terms of the effects that they exhibit. Here, we explore some questions left open by this work.

We do not pursue a completely general study of the most general scenarios involving accelerating branes and brane-bound observers. The chain of transformations we have exploited will not apply in these cases, as these observers will not have been boosted from an inertial frame. However, intermediate steps could be useful in shedding light on possible routes, such as considering an accelerating observer on a Lorentz boosted brane. Ultimately, a tractable approach for general combinations of brane and observer motions would be of interest. 

Our analysis is purely kinematic, neglecting any interactions between branes, observers, and signals. What might happen when dynamics are properly accounted for? We recall that in both the constant velocity and accelerating cases, signals in opposite directions are redshifted and blueshifted. Blueshifted signals will collide with the brane head-on with a higher energy and more frequently than redshifted signals. If these signals interact with the observer and the brane, it is reasonable to expect that the system will slow down when interacting with blueshifted signals and speed up (but less so) when interacting with redshifted signals. Might this provide a natural braking mechanism? Is it possible that under generic initial conditions, branes that interact with fields that propagate in the extra dimension tend to approach the preferred frame? Developing and exploring simple models of such dynamically interacting brane scenarios would be worthwhile, potentially shedding light on how these models might be applied in a cosmological setting.

As mentioned in \cite{PhysRevD.106.085001}, exploring extensions to more general compactifications could help bring us closer to developing more realistic brane-world cosmological models. For example, the authors of  \cite{greene2025compactificationorientationtopologicalscenario,greene2025kleinbottlecosmology} study a novel mechanism for CP violation and baryogenesis involving a brane moving on a spacetime whose extra dimensions are compactified on a Klein Bottle. The work describes the brane as slowing down as particles are produced, raising interesting questions about how our analysis of observers on accelerating branes might be suitably extended and applied to that scenario.

We have also left for future work questions around quantum field theory in the context of moving branes and compact dimensions, building on the analysis found in \cite{kabat2023inducedlorentzviolationmoving}.  What apparent quantum effects might an observer see when their brane is accelerating? Might a quantum field theoretic approach help shed light on the ultimate dynamics of branes in the extra dimension?

\section*{Acknowledgements}

The authors thank Saswat Sarangi for valuable discussions that inspired this work. We thank Brian Greene and Daniel Kabat for helpful comments.

\bibliography{apssamp}

\end{document}